\documentclass[onecolumn]{emulateapj}

\usepackage{hyperref}

\RequirePackage{natbib}
\def\figver{1}
\def\yesfig{1}

\newcommand{\dofig}[1]{
\ifx\figver\yesfig
{#1}
\fi
}

\def\yt{\texttt{yt}}
\def\2HOT{\texttt{2HOT}}

\begin{document}

\title{Dark Sky Simulations: Early Data Release}

\author{Samuel~W.~Skillman\altaffilmark{1}, 
  Michael~S.~Warren\altaffilmark{2},
  Matthew~J.~Turk\altaffilmark{3},\\ 
  Risa~H.~Wechsler\altaffilmark{1,4,5},
  Daniel~E.~Holz\altaffilmark{6},
  P.~M.~Sutter\altaffilmark{7,8,9}
}

\altaffiltext{1}{Kavli Institute for Particle Astrophysics and Cosmology, P.O. Box 2450, Stanford, CA 94305, USA}
\altaffiltext{2}{Theoretical Division, LANL, Los Alamos, NM 87545}
\altaffiltext{3}{NCSA, University of Illinois, Urbana-Champaign, IL 61820}
\altaffiltext{4}{Department of Physics, Stanford University, 382 Via Pueblo Mall, Stanford, CA 94305, USA}
\altaffiltext{5}{SLAC National Accelerator Laboratory, 2575 Sand Hill Road, Menlo Park, CA 94025, USA}
\altaffiltext{6}{Enrico Fermi Institute, Department of Physics, and KICP, University of Chicago, Chicago, IL 60637}
\altaffiltext{7}{Sorbonne Universites, UPMC Univ Paris 06, UMR7095, Institut d’Astrophysique de Paris, F-75014, Paris, France}
\altaffiltext{8}{CNRS, UMR7095, Institut d’Astrophysique de Paris, F-75014, Paris, France}
\altaffiltext{9}{Center for Cosmology and AstroParticle Physics, Ohio State University, Columbus, OH 43210}

\email{samskillman@gmail.com}

\submitted{\texttt{1.4-1-85145466fc56
}\today}

\begin{abstract}

  The Dark Sky Simulations are an ongoing series of cosmological
  N-body simulations designed to provide a quantitative and accessible
  model of the evolution of the large-scale Universe.  Such models are
  essential for many aspects of the study of dark matter and dark
  energy, since we lack a sufficiently accurate analytic model of
  non-linear gravitational clustering. In July 2014, we made available to the
  general community our early data release, consisting of over 55
  Terabytes of simulation data products, including our largest
  simulation to date, which used $1.07 \times 10^{12}~(10240^3)$
  particles in a volume $8h^{-1}\mathrm{Gpc}$ across.  Our simulations
  were performed with \2HOT, a purely tree-based adaptive N-body
  method, running on 200,000 processors of the Titan supercomputer,
  with data analysis enabled by \yt.  We provide an overview of the
  derived halo catalogs, mass function, power spectra and light cone
  data. We show self-consistency in the mass function and mass power
  spectrum at the 1\% level over a range of more than 1000 in particle
  mass.  We also present a novel method to distribute and access very
  large datasets, based on an abstraction of the World Wide Web (WWW)
  as a file system, remote memory-mapped file access semantics, and a
  space-filling curve index.  This method has been implemented for our
  data release, and provides a means to not only query stored results
  such as halo catalogs, but also to design and deploy new analysis
  techniques on large distributed datasets.

\end{abstract}

\setlength{\bibsep}{0.0pt}
\vspace{1em}
\noindent

\maketitle

\section{Introduction}

\begin{figure*}[htb]
\resizebox{1.0\textwidth}{!}{\includegraphics{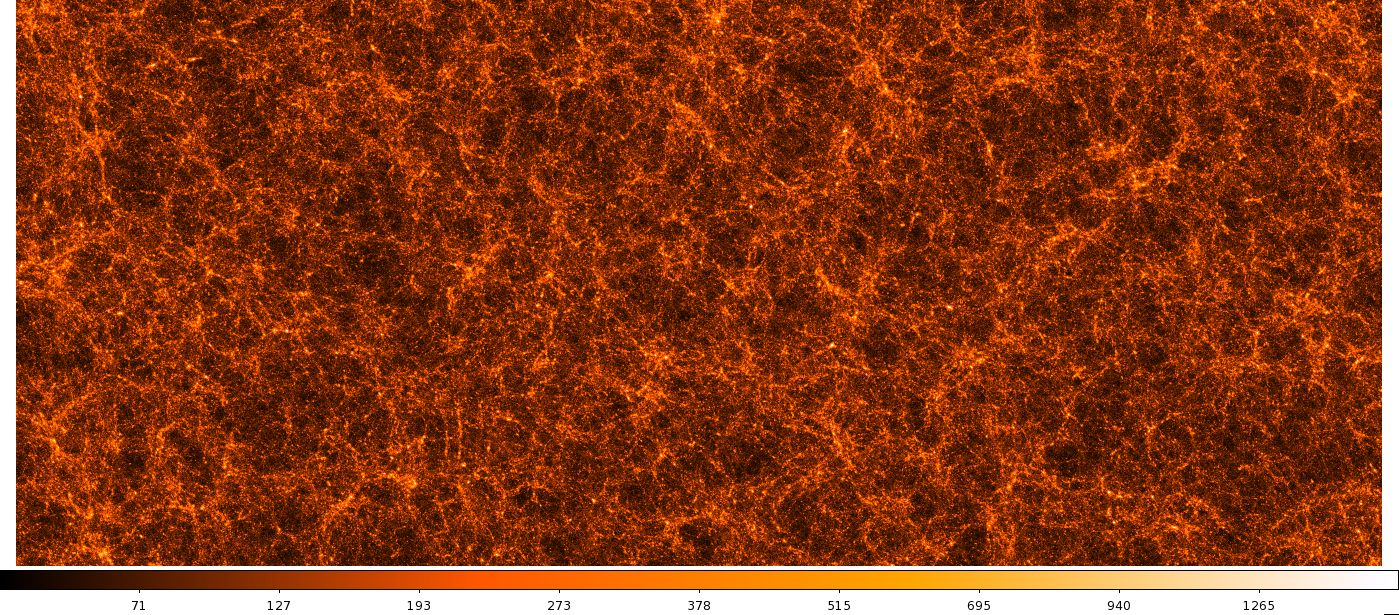}}
\caption{An image of the projected mass density in the $10240^3$ particle ds14\_a simulation lightcone between 
redshift $z=0.9$ and $z=1$.  The angular extent of the image is about 30 $\times$ 15 degrees.  The
color scale represents the number of particles in a HEALPix$^8$~\citep{gorski05} 
pixel with $N_{side}=4096$ (51.5 arcsecond pixels).  This region represents about 1/10,000 of the total simulation volume.}
\label{fig:ds9_slice}
\end{figure*}

In the past 40 years we have witnessed tremendous growth in the
availability and utility of computational techniques and resources.
This has led to an equally astounding increase in the quality and
accuracy of N-body simulations of cosmological structure formation,
starting with $\sim1000$ particles in early works to the current work
(Figure~\ref{fig:ds9_slice})
with 9 orders of magnitude more particles
\citep{peebles70, press74, davis85, efstathiou85, efstathiou90,
jenkins98, klypin99, springel05, springel05a, 2011ApJ...740..102K,
angulo12, alimi12}. N-body simulations form a major pillar
of our understanding of the cosmological standard model; they are an essential link in the
chain which connects particle physics to cosmology, and similarly
between the first few seconds of the Universe to its current state.
Predictions from numerical models are now critical to almost every
aspect of precision studies of dark matter and dark energy, due to the
intrinsic non-linearity of the gravitational evolution of matter in
the Universe. Current and upcoming optical surveys that probe baryon
acoustic oscillations (BAO), the galaxy power spectrum, weak gravitational lensing,
and the abundances of galaxy clusters all require support from
numerical simulations to guide observational campaigns and interpret
their results~\citep{1996ApJ...471...30H, 2005ApJ...633..560E,
2013arXiv1309.5382K, 2013PhR...530...87W, 2014MNRAS.441...24A,
2008PhR...462...67M, 2013arXiv1309.5385H}. Surveys in the X-ray and
sub-millimeter
wavelengths provide an alternate view of the high-temperature plasma
that sits in the deep gravitational potential wells of the dark
matter, and provide an alternate constraint on the abundances of the
most massive collapsed objects in the Universe~\citep{2013arXiv1303.5080P,
2014MNRAS.440.2077M}.

Existing and future observational data motivate our work to understand
theoretically a variety of spatial scales.  The next generation of
surveys will span very large volumes; for example, the Dark Energy
Spectroscopic Instrument (DESI)~\citep{2013arXiv1308.0847L} will survey $\sim$ 30 million
galaxies and quasars over 14000 sq. degrees beyond $z \sim 2.3$,
spanning a volume of $\sim$ 50 (Gpc $h^{-1}$)$^3$.  The Large Synoptic
Survey Telescope (LSST)~\citep{ivezic08} will survey half the sky, detecting
$L^*$ galaxies over a volume of roughly 100 (Gpc $h^{-1}$)$^3$.  Planck
is already able to identify massive galaxy clusters over the entire
sky, beyond $z \sim 1$~\citep{2013arXiv1303.5089P}.  Achieving the science goals of these
surveys requires realistic mock catalogs based on numerical
simulations that calculate the non-linear evolution of structure and
predict the dependence of survey observables on cosmological
parameters.  They also require that these simulations be of
sufficently large volumes to calculate the statistics of rare objects
and to calculate covariances between observables---this requires
multiple realizations as large as survey volumes.

On the largest scales, the universe is populated by clusters of galaxies, connected by filaments,
bordering cosmic voids.  The statistics of the distribution of these
structures can be used in a variety of methods to constrain
fundamental cosmological parameters.  For example, the number of
objects in the Universe of a given mass, the mass function, is
sensitive to cosmological parameters such as the matter density,
$\Omega_m$, the initial power spectrum of density fluctuations, and
the dark energy equation of state.  Especially for very massive
clusters (above $10^{15}$ solar masses [$M_\odot/h$]) the mass
function is a sensitive probe of cosmology.  For these reasons, the
mass function is a major target of current observational programs.
Precisely modeling the mass function at these scales is an enormous
challenge for numerical simulations, since both statistical and
systematic errors conspire to prevent the emergence of an accurate
theoretical model (see~\cite{reed13} and references therein).  The
dynamic range in mass and convergence tests necessary to model
systematic errors requires multiple simulations at different
resolutions, since even a $10^{12}$ particle simulation does not have
sufficient statistical power by itself.

While galaxies and clusters of galaxies account for large
concentrations of mass, cosmic voids that grow from regions of local
divergence are the underdense regions that comprise most of the volume
in the Universe~\citep{2014MNRAS.438.3177S}.  Today, over two thousand
voids have been detected in galaxy redshift surveys
\url{http://www.cosmicvoids.net} and they offer excellent probes of
cosmology via their size distributions, shapes, internal dynamics,
and correlations with the Cosmic Microwave Background, as well as
unique probes of magnetic fields and galaxy evolution.  Voids are only
observed in the galaxy distributions, and galaxies are sparse, biased
tracers of the underlying dark matter.  However, voids are typically
studied from a theoretical perspective only in dark matter $N$-body
simulations.  The identification of voids is sensitive to survey density and
geometry in a highly non-trivial fashion; to make direct contact with observed
voids we must perform large-volume, high-resolution simulations to
capture the structure and dynamics of dark matter, map the dark matter
to a galaxy population, place the galaxies on a lightcone, apply
realistic survey masks, and identify and characterize the voids.

\footnotetext{\url{http://healpix.jpl.nasa.gov}}

It is theorized that small perturbations, referred to as baryon
acoustic oscillations (BAO) and possibly excited during an
inflationary epoch, launched sound waves in the photon-dominated
baryon plasma. As the Universe expanded and the plasma cooled,
eventually these perturbations were ``frozen-in'' at the time of
recombination, and are seen as the fluctuations in the Cosmic
Microwave Background~\citep{bennett12, planckcollaboration13, 2013arXiv1308.0847L}.  These small
fluctuations are thought to lead to an imprint in the spatial
distribution of large scale structure, which can be measured directly
by a number of galaxy surveys~\citep{2005ApJ...633..560E,
2014MNRAS.441...24A, 2011MNRAS.416.3017B, 2013arXiv1308.0847L} and in
upcoming low-frequency radio surveys~\citep{johnston08, dewdney09}.  The
BAO signal has been detected at $\sim10\sigma$ in the Sloan Digital
Sky Survey (SDSS-III) Data Release 11 (DR11) Baryon Oscillation Spectroscopic 
Survey (BOSS) galaxy samples~\citep{2014MNRAS.441...24A}.
 In principle the precise
structure of the galaxy distribution can be used to probe
cosmological parameters.  Our theoretical models need to keep up with
the tremendous advances in observational data, and high quality dark
matter simulations can be used provide a bridge between observational
and theoretical cosmology.  The most basic statistical measures of
galaxy clustering are the power spectrum and the two-point correlation
function. By producing high-quality databases of galaxy tracers
(i.e.~``mock catalogs''), cosmological simulations are able to probe
observed galaxy distributions~\citep{2000MNRAS.318.1144P,
2006ApJ...652...71W, 2012ApJ...745...16T, 2014ApJ...783..118R}.
Galaxy velocities can also be used for directly testing
cosmology~\citep{johnston12}.

The rigorous statistical and systematic demands of upcoming surveys
requires the computational cosmology community to design and deliver
high quality simulations that can be used to further our understanding
of cosmological theory and the large scale structure of our universe.
Our measurements of the Universe are sufficiently refined such that we
now require both large statistical volumes and accurate,
high-performance methods.  \cite{kuhlen12} reviews the prior
state-of-the-art in numerical simulations of the Dark Universe, the
largest being the ``DEUS FUR'' 550 billion particle
simulation~\citep{alimi12} performed with the \verb-RAMSES- adaptive
particle-mesh code~\citep{teyssier02}.  Other simulations/codes at the
$10^{11}$ particle scale are HR3~\citep{kim11} using
\verb-GOTPM-~\citep{dubinski04}, Millennium-XXL~\citep{angulo12} with
\verb-GADGET3-~\citep{springel05}, Jubilee~\citep{watson13} with
\verb-CUBEP3M-~\citep{harnois-deraps12}, and
Bolshoi/BolshoiP~\cite{2011ApJ...740..102K} with \verb-ART-~\citep{kravtsov97}.
Other codes that have advertised capability at the $10^{12}$ particle
scale are \verb-HACC-~\citep{habib13} and Gordon Bell prize winner
\verb-Greem-~\citep{ishiyama12}.  The method most commonly used for
accessing these simulations are relational databases which allow SQL
queries to halo catalogs or other data from the simulations
\citep{lemson06,riebe13}.

A potentially disturbing observation is that the research cycle
associated with cutting-edge simulations is dominated not by the
runtime of the simulation itself, but in the time to both extract
scientific results and (if at all) make the simulation data publicly
available.  This indicates that porting existing analysis tools,
validating results, and developing new analysis techniques have not
received the same attention as our primary simulation codes. 
This is not a new phenomenon; in our own work for the 1992 Gordon Bell prize~\citep{warren92}, it
took another two years for scientific results to become
available~\citep{zurek94}.  Recent examples of a similar timescale are
the Millennium-XXL simulation completed in Summer 2010 with results
submitted in March, 2012~\citep{angulo12}, and the DEUS FUR simulation
completed in March 2012, with results in~\cite{rasera14}.  This highlights
the fact that software development and data-analysis are less amenable
to acceleration from advances in computer hardware, and warrants
additional attention regarding the most productive allocation of resources
to support sustainable scientific software and simulations.

Computational techniques have progressed so rapidly that the time to
run a state-of-the-art simulation following $13.8$ billion years of
cosmic history takes only days on modern supercomputer architectures.
However, this progress has not come without a cost. The time to
design, develop, debug, and deploy a simulation can take years. After
the completion of a simulation, the time to disseminate the main
results takes months or years. Public data releases can take years to
happen, if at all.  There are many reasons for the significant delays
in dissemination, both technical and social. Simulations of these
types produce raw datasets that are measured in hundreds of Terabytes
or even Petabytes, with even reduced halo catalogs reaching Terabytes
in size.  Even on high speed networks, data transfer of just a single
snapshot can take as long as the original simulation.  Socially, there
is both the concern that a mistake will be discovered and the worry
that others will make important discoveries with the publicly released
data, thereby curtailing the scientific accomplishments of the
simulators.  However, despite these concerns, we believe it is of
immense value to the community for everyone to release their
simulation data as soon as possible.  A mistake found by another can
be fixed, and a new iteration of a simulation can be undertaken.  A
discovery made by outsiders is only possible through the hard work of
the data creators, and correct attribution makes that known.

Open source software projects have led the open data field, as 
a rapid increase in the availability of software developed in the open has
led to burgeoning communities of developers in many astrophysical projects
(\texttt{yt}, \texttt{astropy}, and \texttt{sunpy} to only name a few of the
Python-based projects).
A primary goal of this project is to adopt some of the fundamental
concepts of the open source community, and translate them to open data for state
of the art cosmological simulations.  We aim to decrease the barrier to entry
for accessing and analyzing data, and increase the speed of iteration and 
pace of scientific inquiry.  Through a set of ongoing simulations, we begin this progress by releasing
both reduced and raw data products from a set of cosmological simulations, including a
simulation utilizing over a trillion particles.  We hope to use input and
feedback from the broader community to help shape our future data releases,
and ultimate the types of simulations to be run. Our hope is that these
data products enable discovery pertaining to many areas of research in
cosmology.  In particular, our trillion particle simulation is included in our
current release; this is state-of-the-art in terms of mass resolution for its
cosmological volume, and should offer tremendous insight into the largest scales
and structures in our Universe. 

While experts can be trained to interact with large datasets on parallel file
systems, we aimed to create an interface to the data that is novel, simple, and
extensible, with the aims of allowing people with a wide range of technical abilities and
interests to explore the data.  In principle, high school students
interested in physics and/or computation should be capable of accessing subsets
of a trillion particle dataset and studying the structure of the dark matter
potential in a sample galaxy cluster.  Researchers in large scale data
visualization should be able to load halo catalogs into 3D models of our
universe, and explore alternate representation methods for high-dimensional
datasets.  Digital artists or game designers may even be inclined to use our
data as input for their personal work.

In what follows we describe the simulation setup, data analysis
pipeline, and data access methods. We also describe our initial data
validation through the analysis of the $z=0$ mass function, power
spectra, and a brief comparison to the Planck Sunyaev-Zel'dovich
galaxy cluster catalog.  We end with a proposed set of community
standards for fostering growth in computational cosmology both within
and exterior to the confines of the Dark Sky Simulations project.

\vspace{5mm}
\section{Software \& Hardware}
\label{sec:Methods}
\subsection{\2HOT}
\label{sub:2hot}
\2HOT is an adaptive treecode N-body method whose operation count
scales as $N \log N$ in the number of particles.  It is described
in~\cite{warren13}, which we summarize here, and offer additional
details relevant to the Dark Sky Simulations Early Data Release.  Almost 30
years ago, the field of N-body simulations was revolutionized by the
introduction of methods which allow
N-body simulations
to be performed on arbitrary collections of bodies in a time much less
than $O(N^2)$, without imposition of a lattice
\citep{appel85,barnes86a,greengard87}.  They have in common the use of
a truncated expansion to approximate the contribution of many bodies
with a single interaction.  The resulting complexity is usually
determined to be $O(N)$ or $O(N \log N)$, which allows computations
using orders of magnitude more particles.  These methods represent a
system of $N$ bodies in a hierarchical manner by the use of a spatial
tree data structure.  Aggregations of bodies at various levels of
detail form the internal nodes of the tree (cells).  These methods
obtain greatly increased efficiency by approximating the forces on
particles. Properly used, these methods do not contribute
significantly to the total solution error.  This is because the force
errors are exceeded by or are comparable to the time integration error
and discretization error.  Treecodes offer the best computational
efficiency when force resolution at small scales is important.  \2HOT
is distinguished from other current cosmology simulation approaches at
the petascale by not having a particle-mesh component, using a pure
treecode avoiding the potentially problematic transition scale between
PM and tree forces inherent with TreePM approaches, and offering
additional flexibility for high-resolution simulations with a large
dynamic range in particle masses.

The code has been evolving for over 20 years on many computational
platforms.  We began with the very earliest generations of
distributed-memory message-passing parallel machines (an architecture
which now dominates the arena of high-performance computing), the
Intel iPSC/860, Ncube machines, and the Caltech/JPL Mark\,III
\citep{warren88,warren92a}.  This original version of the code
was abandoned after it won a Gordon Bell Performance Prize in 1992
\citep{warren92}, due to various flaws inherent in the code, which was
ported from a serial version.  A new version of the code was initially
described in~\cite{warren93}.  Since then, our hashed oct-tree (\texttt{HOT})
algorithm has been extended and optimized to be applicable to more
general problems such as incompressible fluid flow with the vortex
particle method~\citep{ploumhans02} and astrophysical gas dynamics with
smoothed particle hydrodynamics (SPH)~\citep{fryer06}.  The code also won the Gordon Bell
performance prize again in 1997, with absolute performance reaching
430 Gflops on ASCI Red on a 320 million particle simulation, as well
as obtaining a Gordon Bell price/performance prize on the Loki Beowulf
cluster~\citep{warren97a} and the Avalon Beowulf cluster~\citep{warren98}.

The basic algorithm may be divided into several stages.  First, particles are domain
decomposed into spatial groups.  Second, a distributed tree data
structure is constructed.  In the main stage of the algorithm, this
tree is traversed independently in each processor, with requests for
non-local data being generated as needed.  In our implementation, we
assign a \verb-Key- to each particle, which is based on Morton
ordering~\citep{samet90}.  This maps the points in 3-dimensional space to a
1-dimensional list, while maintaining as much spatial locality as
possible.  The domain decomposition is obtained by splitting this list
into pieces.  
The Morton ordered key labeling scheme implicitly defines the topology
of the tree, and makes it possible to easily compute the key of a
parent, daughter, or boundary cell for a given key.  A hash table is
used in order to translate the key into a pointer to the location
where the cell data are stored.  This level of indirection through a
hash table can also be used to catch accesses to non-local data, and
allows us to request and receive data from other processors using the
global key name space.  We have developed an efficient mechanism for
latency hiding in the tree traversal phase of the algorithm, which is
critical to high performance.

A recent major effort on code development has added many additional
features to the code, being designated in the naming transition from
\texttt{HOT} to \2HOT.  Accuracy and error behavior have been improved
significantly for cosmological volumes through the use of a technique
to subtract the uniform background density~\citep{warren13},
correcting for small-scale discretization error, and using a Dehnen
$K1$ compensating smoothing kernel~\citep{dehnen01} for small-scale
force softening.  We use an adaptive symplectic
integrator~\citep{quinn97} and an efficient implementation of periodic
boundary conditions using a high-order ($p=8$) multipole local
expansion~\citep{challacombe97,metchnik09} which accounts for the
periodic boundary effects to near single-precision floating point
accuracy (one part in $10^{-7}$).  We adjust the error tolerance
parameter to limit absolute errors to 0.1\% of the rms peculiar
acceleration.  Our code and parameters have been extensively tested
and refined with thousands of simulations to test accuracy and
convergence across multiple dimensions of timestep, smoothing length,
smoothing type, error tolerance parameters, and mass resolution.

The \2HOT code is written in the \verb-C- programming language.  We utilize
a variety of \verb-gcc- extensions, the most important of which is the
\verb-vector_size- attribute, which directs the compiler to use
\verb-SSE- or \verb-AVX- vector instructions on Intel architectures.
Using gcc with \verb-vector_size- has eliminated the need to write the
gravitational inner loops in assembly language to obtain good
performance on CPU-only architectures.  We have implemented the GPU
portions of our code in both \verb-CUDA- and \verb-OpenCL-, with the
\verb-CUDA- versions performing somewhat better at present.  We use a
purely message-passing programming model, implemented in \verb-MPI-.
In order to hide latency, the tree-traversal phase of our algorithm
uses an active message abstraction implemented inside \verb-MPI- with
our own ``asynchronous batched messages'' routines.  Our \2HOT
software does not depend on any external libraries.

Treecodes place heavy demands on the various subsystems of modern
parallel computers.  This results in very poor performance for
algorithms which have been designed without careful consideration of
message latency, memory bandwidth, instruction-level parallelism and
the limitations inherent in deep memory hierarchies.
The \emph{space-filling curve domain decomposition} approach we proposed
in~\cite{warren93} has been widely adopted in both application codes
(e.g. \cite{griebel99,fryxell00,springel05,gittings08,jetley08,wu12}) and more general
libraries~\citep{parashar96,macneice00}.  Our claim that such orderings
are also beneficial for improving memory hierarchy performance has
also been validated~\citep{mellor-crummey99,springel05}.
Our method converts a $d$-dimensional set of data elements
into a 1-dimensional list, while maintaining as much spatial locality
in the list as possible.  This allows us to neatly domain decompose
any set of spatial data.  The idea is simply to cut the
one-dimensional list of sorted elements into $N_p$ (number of
processors) equal pieces, weighted by the amount of work corresponding
to each element. The implementation of the domain decomposition is
practically identical to a parallel sorting algorithm, with the
modification that the amount of data that ends up in each processor is
weighted by the work associated with each item.  The mapping of
spatial co-ordinates to integer keys converts the domain decomposition
problem into a generalized parallel sort.  The method we use is
similar to the sample sort described in~\cite{solomonik10}. Note
that after the initial decomposition, the \verb-Alltoall-
communication pattern is very sparse, since usually elements will only
move to a small number of neighboring domains during a timestep.  This
also allows significant optimization of the sample sort, since the
samples can be well-placed with respect to the splits in the previous
decomposition.  In~\cite{warren95a} we describe a tree traversal
abstraction which enables a variety of interactions to be expressed
between ``source'' and ``sink'' nodes in tree data structures.  This
abstraction has since been termed \emph{dual-tree traversal}~\citep{yokota12}.
The dual-tree traversal is a key component of our initial
approach to increase the instruction-level parallelism in the code to
better enable GPU architectures.  It is also relevant to a number of
data analysis tasks, such as neighbor-finding and computing
correlation functions.

In this earlier work we used the fact that particles which are
spatially near each other tend to have very similar cell interaction
lists.  By updating the particles in an order which takes advantage of
their spatial proximity, we improved the performance of the memory
hierarchy.  Going beyond this optimization with dual-tree traversal,
we can bundle a set of $m$ source cells which have interactions in
common with a set of $n$ sink particles (contained within a sink
cell), and perform the full $m \times n$ interactions on this block.
This further improves cache behavior on CPU architectures, and enables
a simple way for GPU co-processors to provide reasonable speedup, even
in the face of limited peripheral bus bandwidth.  We can further
perform data reorganization on the source cells (such as swizzling
from an array-of-structures to a structure-of-arrays for SIMD
processors) to improve performance, and have this cost shared among
the $n$ sinks.  In an $m \times n$ interaction scheme, the interaction
vector for a single sink is computed in several stages, which requires
writing the intermediate results back to memory multiple times.
For current architectures, the write bandwidth available is easily
sufficient to support the $m \times n$ blocking.

This is the key to our approach on Titan enabling high performance 
from the GPUs: we bundle multiple particles
with a single interaction list to increase the computational
intensity. This allows the full GPU performance to be sustained
without being severely limited from PCI-Express bandwidth.  In more
detail, the computational intensity of our inner loop is 2 flops per
byte.  With an achievable PCI bandwidth of 5 Gbytes/sec on Titan, we
need to increase the flops/byte by a factor of 200 to support
a 2 Tflop GPU.  We achieve this by packaging 200 or more particles
with the same interaction list and sending them to the GPU.  Finding
more than 200 particles which share the same cell interactions is only
possible given the framework of the \2HOT code, which provides the
grouping and multipole acceptance tests to arrange the computation
suitably.  Even then, this technique only works for about 80\% of the
interactions, with the rest near the leaf nodes of the tree not being
able to be sufficiently grouped.

A second round of GPU optimizations was required to manage most of the
remaining interactions near the leaf nodes.  In particular,
pre-staging a large block of particle positions and terminating the
tree traversal once less than 80 particles remained in a cell
(handling the rest with direct interactions) and bundling partial
lists for leaf level quadrupole and hexadecapole
interactions allows us to perform 97\% of the gravitational
interactions on the GPU, with all of the tree traversal logic handled
by the CPU.

As a scaling comparison (see Table~\ref{tab:one_node}), we have run
the \2HOT code on the same small cosmology problem using a single node
of the Titan supercomputer, as well as Eos (an Intel Xeon E5-2670
processor-based Cray XC30 also at Oak Ridge National Laboratory) a
desktop Haswell processor, and an Amazon Elastic Compute Cloud (EC2)
current generation {\tt c3.8xlarge} and older {\tt c1.large} instance.
We quote performance in particles updated per second per node (p/s/n),
where our 5.9 Petaflop result described in Section~\ref{sub:ds14_a} below
corresponds to $10240^3/110/12288 = 7.94 \times 10^5$ p/s/n, or 64\%
efficiency scaling from 1 node to 12288 nodes.

The most scientifically relevant metric for evaluating gravitational
N-body simulations is not Petaflops, but how many particles are
updated per second, at an accuracy sufficient to accurately represent
the physics involved.  In 1997, we obtained a performance of 3 million
particles updated per second at an RMS force accuracy better than
$10^{-3}$~\citep{warren97a}.  Our current performance results are about 8 billion
particles per second, with an equivalent force accuracy about 10 times
better.  Whether this accuracy is sufficient, or if accuracy can be
sacrificed without adversely affecting the scientific results, is an
area of current research.

\begin{table}
\centering
\begin{tabular}{|l|c|c|r|}
\hline  Node Description & Cores & (MHz) & Perf.~(p/s/n) \\  \hline
Opteron 6274/K20x & 30 & 2200 & $12.25 \cdot 10^5$ \\
Opteron 6274      & 16 & 2200 & $ 2.54 \cdot 10^5$ \\
Xeon E5-2670 (HT) & 32 & 2600 & $ 6.37 \cdot 10^5$ \\
Xeon E5-2670      & 16 & 2600 & $ 5.78 \cdot 10^5$ \\
EC2 c3.8xlarge    & 32 & 2800 & $ 3.91 \cdot 10^5$ \\ 
EC2 c1.xlarge     &  8 & 1800 & $ 1.00 \cdot 10^5$ \\ 
Core i5-4570      &  4 & 3200 & $ 2.50 \cdot 10^5$ \\
\hline
\end{tabular}
\label{tab:one_node}
\caption{Performance measured in particles updated per second per node
  (p/s/n) for a variety of computational platforms.  The top line is a
  Titan node (including the NVIDIA K20x GPU counted as 14 cores),
  which is 4.8x faster than the second line (without the
  GPU).  The third and fourth lines compare a Cray XC30 node with and
  without hyperthreading.  The Amazon EC2 results had a well defined
  price, which was \$0.263 per hour for the c3.8xlarge instance, and
  \$0.064 for the c1.xlarge instance.  Performing an equivalent number
  of particle updates as our large ds14\_a run using Amazon EC2
  resources would have cost at least \$200,000 at the current Amazon
  spot price, or \$1.3 million for on-demand (\$1.68/hr) c3.8xlarge resources.}
\end{table}

\subsection{\texttt{SDF}}
\label{sub:SDF}

We use the Self-Describing File (SDF) interface, originally
designed and implemented for our early parallel simulations~\citep{warren92}, with an
implementation of the interface recently released under an open-source license~\citep{sdf2014}. The basic aims
of the SDF library are to be simple, flexible, extensible and most importantly, scalable
to millions of processing elements.  By writing analysis software which uses the SDF interface, 
the differences between data formats can be encapsulated, allowing software to read multiple
data layouts and formats, without requiring recompilation each time a field is added, or a new data format
needs to be supported.
The SDF format consists
of a human readable ASCII header followed by raw binary data. The header is
intended to support metadata that describes the data, its pedigree,
checksums of the data contained within it, and its layout on disk or in memory.  
The header also provides all the information needed for any
processor in a parallel machine to independently read its own portion of a dataset.
Note that the interface is capable of describing other existing data formats (such
as the GADGET and Tipsy formats commonly used for cosmological datasets).  In this regard,
it is distinguished from libraries such as HDF5~\citep{folk99}, which can not describe existing data without
going through a conversion process.
Each line in the header that includes an \texttt{=}
is interpreted as a key-value dictionary pair. Other lines are
interpreted as comments or internal SDF parameters.  The structure of
the binary data is encoded using a structure descriptor closely analogous to
a C language structure declaration.  One is
allowed to have as many fixed-length arrays as desired followed by an
optional arbitrary length array as the last entry (whose length can be determined by the structure size and
file length).  We note that
while the majority of the data presented in this EDR follows the
array-of-structures (AoS) layout, structure-of-array (SoA) is also
enabled by the SDF format.  
This data format is portable and we provide both a C and Python
library for reading the headers and binary data.  We
also provide a Python interface for specific spatial queries such as
bounding boxes and spheres through \texttt{yt}, described in
Section~\ref{sub:yt}. In addition, frontends for the raw particle and
halo catalog SDF datasets have been added to \texttt{yt}.  

The majority of our data is stored in spatial Morton order (also known as \texttt{Z} or \text{N} ordering), meaning
that the sequential particles or halos on disk lie along a space filling
curve. We further expose this embedded spatial structure in the file by
creating an auxiliary file, which
we refer to as the Morton index \texttt{midx} file. This \texttt{midx}
file is constructed by first choosing a level in the oct-tree hierarchy
to bin particles.  The full particle dataset is then searched to find
the first particle offset and the total number
of particles within each leaf node in the tree.  These values are mapped to the
Morton index \verb~Key~.  This file has an
extension with the naming convention \texttt{.midx\%i \% level}, where
\texttt{level} corresponds to the level of the oct-tree that was
constructed to bin the particles.  For example, ``.midx7'' corresponds
to a Morton index file that bins particles into $(2^7)^3$ cells. The
\texttt{midx} file itself is also stored in the SDF format. We typically
use higher level \texttt{midx} files for progressively larger data.

Because the size of an individual \texttt{SDF} file can be many Terabytes,
we have utilized the concept of a memory-mapped file in two separate
implementations. The first is exposed through the Python interface using
a \texttt{Numpy} \texttt{memmap}. This creates an ``out-of-core'' 
view of the complete data file.  We have utilized this technique for the
majority of our early science results, and have been satisfied with its
performance and ease-of-use.

Our second implementation extends this concept to file stored on the
World Wide Web (WWW), and utilizes a local page-cache mechanism to
address a remote file through a thin wrapper that exposes binary data
to \texttt{Numpy}.  This approach transparently takes advantage of the
enormous investment in multiple technologies which have been developed to improve the
performance and reliability of generic WWW resources.
We released this software at the time of this
manuscript's submission~\citep{Turk:10773}.  This is the first instance
that we are aware of that directly exposes binary data hosted on the
WWW into local memory in a running Python session.  We note that this
interface may be useful for future large astronomical surveys, as we
will later describe how we use this technology to address individual
files that are 34 TB\footnote{$1 \mathrm{TB} \equiv 10^{12}
\mathrm{bytes}$} in size, similar to the expected size of individual
data products from SKA-1 survey in
2020~\citep{2014arXiv1403.2801K}.

\subsection{\yt}
\label{sub:yt}

From the data perspective, \2HOT can be thought of as a highly efficient
method to create vast amounts of unstructured data.  We therefore required
an analysis framework that is both capable of ingesting Terabytes of data
and allows for rapid design and development phases for analysis.  For this
reason we have utilized and extended \yt, an open-source analysis and
visualization package written primarily in \verb~Python~, which our team has
experience in applying to both large unigrid simulations (such as the $3600^3$
radiation-hydrodynamics simulation mentioned in~\cite{2013arXiv1306.0645N}) and
deep Adaptive Mesh Refinement simulations of the first stars~\citep{2009Sci...325..601T}.
\yt~is parallelized using mpi4py~\citep{dalcin08}, which
we have exposed on $2048$ nodes ($32768$ cores) of the Titan supercomputer in
this work.

\texttt{yt}~was originally designed to manage data output by patch-based
Adaptive Mesh Refinement (AMR) astrophysical simulations.  Recent versions have
restructured the underlying engine to shift the focus from AMR simulations to
other forms of data such as octree, unstructured mesh, and as used here,
particle-based datasets.  During this transition, the focus of \texttt{yt} has
shifted to enable faster and more flexible indexing methods, which are utilized
here to great extent, ensuring that even the very largest of datasets can be
analyzed with proper care taken to enable multi-level indexing and on-demand
data loading.  

In this work we have used \yt~as a base upon which we build access methods to
both local and remote (on the WWW via \texttt{thingking}) Petascale
datasets.  We required a system that enabled ease of deployment, ease of
access, and minimized the burden on the
individual researcher interested in examining the data --- and perhaps most
importantly, we rejected any solution that obfuscated the data in any way.  For
this reason, rather than presenting a SQL frontend, or a science gateway for
exploring pre-selected data products and results, we have instead taken
a hybrid approach that preserves access to the underlying data, while still
making accessible the reduced data products.  Our extensions include adding
an \texttt{SDF} frontend that can utilize the \texttt{midx} index for spatially
querying large datasets (up to 34 TB at the time of this writing). These additions
are being actively reviewed through the \yt~projects peer review system, and 
in the meantime can be used through a fork that can be located through our
project website (\url{http://darksky.slac.stanford.edu}).

The current version of \texttt{yt} also provides support for loading halo
catalogs generated by ROCKSTAR as particle datasets, enabling selection,
processing and visualization of the halo objects.  We envision users loading
the halo catalogs, determining the regions of interest to them in the full
dataset, and then (transparently) utilizing the multi-level indexing system
described above to load only the subselection of particles that are relevant
to their research questions.  While it would normally be completely intractable
for a researcher to analyze a 34 TB file of particles, this approach will make
it convenient and straightforward.

\subsection{Halo Finding}
\label{sub:HaloFinding}
To identify dark matter halos and substructure we use the ROCKSTAR
algorithm~\citep{behroozi13}.  This halo-finding approach is based on
adaptive hierarchical refinement of friends-of-friends groups in both
position and velocity.  It has been tested extensively and compared
with other halo finders in~\cite{knebe11}, showing excellent, and in many
cases superior, performance when compared with other approaches.

While we have previously used the basic ROCKSTAR code successfully for
simulations with 69 billion particles using 10,000 CPUs, several
features of the code were not ideal for the computing environment on
the Titan system.  In particular, the client-server model of
computation in ROCKSTAR requires using a file descriptor for each pair
of communicating processes, and the limit on open file descriptors for
each CPU on Titan is 32768.  In addition, we have found it preferable
to use a smaller number of CPUs to process the data in pieces, which
enables the use of less capable computing resources for analysis
(rather than requiring a machine with over 70 Tbytes of RAM to process
the simulation all at once).

In our implementation, we take advantage of the modularity of the
ROCKSTAR algorithm and use only the function interface
\verb-find_subs()- via the \texttt{yt-3.0} halo finding interface.  We
process each spatial domain independently, with \texttt{yt} loading
the spatially indexed particle data for a domain and finding the
initial FOF groups.  We add a buffer region to each domain to contain
any particles from a halo near the domain edge ($6h^{-1}\mathrm{Mpc}$ is sufficient
for halo masses up to $10^{16}h^{-1}M_\odot$).  These groups are then
passed to ROCKSTAR \verb-find_subs()-.  For strict spherical
overdensity masses, \texttt{yt} performs the same post-processing
steps to assign particles which were missed by the FOF group to halos,
and to identify parent/sub-halo relationships.  We have validated our
implementation by comparing it with unmodified ROCKSTAR for smaller
($4096^3$) simulations. For a 1.07 trillion particle dataset, our
complete halo finding process takes about 6 hours on 1024 CPUs.

\subsection{Computational Hardware}

The complete Titan Cray XK7 system~\citep{bland12} at Oak Ridge
National Laboratory contains 18,688 compute nodes, each containing a
16-core 2200 MHz AMD Opteron 6274 with 32 GB of 1600 MHz DDR3 memory,
paired across a PCI-Express 2.0 bus with an NVIDIA Tesla K20x GPU with
6 GB of memory.  The Cray Gemini interconnect~\citep{alverson10}
provides roughly 8 Gbytes/sec of bi-directional bandwidth per node at
the hardware level, with MPI latencies quoted as 1.5 microseconds or
less.  Titan host nodes currently run the SUSE Linux Enterprise Server
11 SP1 (x86\_64) with current kernel 2.6.32.59.  Processing nodes run
Cray's Compute Node Linux, designed to minimize interference between
operating system services and application scalability.

\subsection{Additional Software}

All of our software was compiled with system installed
gcc~\citep{stallman89} version 4.8.2 20131016 (Cray Inc.), with the
addition of NVIDIA's LLVM-based~\citep{lattner04} nvcc V5.5.0 for two
CUDA-specific files.  Significant software dependencies used via the
system modules interface include cray-mpich/6.2.0~\citep{snir98,gropp03,bosilca02},
fftw/3.3.0.3~\citep{frigo98} and gsl/1.16~\citep{galassi07}.

We additionally installed roughly 40 additional packages that were not
available on the system, or were out-of-date including gdb 7.6.90
(system version 7.5.1), Mercurial~\citep{osullivan09} (not available on system), git
1.8.5~\citep{torvalds05,loeliger12} (system version 1.6.0), Python
2.7.6~\citep{vanrossum95} (system version 2.6.8), MPI for
Python~\citep{dalcin08}, Cython~\citep{behnel11},
Numpy~\citep{oliphant06} and matplotlib~\citep{hunter07}.
We have additionally made extensive use of the \texttt{bbcp} 
program~\citep{hanushevsky01} to transfer
hundreds of Tbytes of data between Oak Ridge, LANL, SLAC and NERSC, as well 
as efficiently copy data between local systems.

\section{The Dark Sky Simulations}
\label{sec:DarkSkySimulations}

We have performed a series of
calculations that vary in particle number from $2048^3$ to $10240^3$ and in
comoving cosmological volumes from $100h^{-1}\mathrm{Mpc}$ to $8h^{-1}\mathrm{Gpc}$ on a side, as shown
in Table~\ref{tab:simulations}.  The set of simulations released in this Early 
Data Release (EDR)
are part of a larger effort~\citep{Warren:10777} enabled through a 2014 INCITE\footnote{\url{http://www.doeleadershipcomputing.org/incite-awards/}} computing grant
at Oak Ridge National Laboratory.  Additionally, one of the
small volume boxes was run on the LANL Mustang supercomputer. 
All of these simulations utilize the exact same cosmology with 
$(\Omega_m, \Omega_b, \Omega_\Lambda, h_{100}, \sigma_8) = (0.295, 0.0468,
0.705, 0.688, 0.835)$, detailed in
Table~\ref{tab:cosmological_parameters}.  Initial conditions were calculated
using a modified version of 2LPTic~\citep{crocce06}.  The input power spectrum
was calculated from a million step Markov Chain Monte Carlo
calculation using MontePython~\citep{audren12a} and CLASS, the Cosmic Linear
Anisotropy Solving System~\citep{blas11}.  Observational constraints used as
inputs included Planck~\citep{planckcollaboration13}, BOSS~\citep{delubac14}
and BICEP2~\citep{bicep2collaboration14}. The raw Markov Chain data the input cosmology was based
on is publicly available~\citep{warren14a}.

\begin{table}[htb]
\centering
\begin{tabular}{|l|c|c|c|c|c|c|}
  \hline
  Simulation & $\sqrt[3]{N}$ & L [$h^{-1}\mathrm{Mpc}$] & $M_p$ [$h^{-1}M_\odot$] & $\epsilon~[h^{-1}\mathrm{kpc}]$ & $z_{\mathrm{init}}$ & timesteps\\
  \hline  
  ds14\_a             &  10240 & 8000 & $3.9 \cdot 10^{10}$ & 36.8 &  93 &  543\\ 
  ds14\_g\_1600\_4096 &  4096  & 1600 & $4.9 \cdot 10^{9}$  & 18.4 & 135 &  563\\ 
  ds14\_g\_800\_4096  &  4096  &  800 & $6.1 \cdot 10^{8}$  &  9.2 & 183 &  983\\ 
  ds14\_g\_200\_2048  &  2048  &  200 & $7.6 \cdot 10^{7}$  &  4.6 & 240 & 1835\\ 
  ds14\_g\_100\_2048  &  2048  &  100 & $9.5 \cdot 10^{6}$  &  2.3 & 297 & 3539\\ 
  \hline
\end{tabular}\\[.25in]
\caption{Dark Sky Simulations.  For each simulation, the name, particle count, box size, particle mass, Plummer equivalent softening length, starting redshift
and number of timesteps  are shown above.}
\label{tab:simulations}
\end{table}

\begin{table}[htb]
\centering
\begin{tabular}{|l|c|c|c|c|} 
 \hline 
Param & best-fit & mean$\pm\sigma$ & 95\% lower & 95\% upper \\ \hline 
$100~\omega_{b }$ &$2.201$ & $2.214_{-0.025}^{+0.023}$ & $2.165$ & $2.263$ \\ 
$\omega_{cdm }$ &$0.118$ & $0.1175_{-0.0015}^{+0.0014}$ & $0.1146$ & $0.1204$ \\ 
$H_0$ &$68.46$ & $68.81_{-0.68}^{+0.64}$ & $67.48$ & $70.13$ \\ 
$10^{+9}A_{s }$ &$2.18$ & $2.187_{-0.058}^{+0.052}$ & $2.076$ & $2.296$ \\ 
$n_{s }$ &$0.9688$ & $0.9676_{-0.0054}^{+0.0054}$ & $0.9568$ & $0.9785$ \\ 
$\tau_{reio }$ &$0.08755$ & $0.09062_{-0.013}^{+0.012}$ & $0.06518$ & $0.1156$ \\ 
$r$ &$0.1511$ & $0.1737_{-0.042}^{+0.035}$ & $0.09785$ & $0.2526$ \\ 
$\Omega_{\Lambda }$ &$0.7012$ & $0.7048_{-0.0081}^{+0.0085}$ & $0.688$ & $0.7211$ \\ 
$YHe$ &$0.2477$ & $0.2477_{-0.00011}^{+0.0001}$ & $0.2475$ & $0.2479$ \\ 
$\sigma_8$ &$0.8355$ & $0.8344_{-0.012}^{+0.011}$ & $0.811$ & $0.858$ \\ 
\hline 
 \end{tabular} \\[.25in]
\caption{Cosmological Parameters. See \citet{warren14a} for details.  Our simulations use the
parameters in the mean value column.  The values of any parameters not specified are available in the \texttt{class.ini} file
in the data repository.}
\label{tab:cosmological_parameters}
\end{table}

\subsection{The ds14\_a and ds14\_g Simulations}
\label{sub:ds14_a}

On April 18-19, 2014, we ran the ds14\_a simulation from redshift $z=93$ to $z=0$ with
1,073,741,824,000 ($10240^3$) particles on 12,288 nodes (196,608 CPU cores and 12,288 NVIDIA K20x GPUs) of the Titan
system at Oak Ridge National Laboratory, which represents approximately 2/3 of the
total machine.  The simulation was of a cubical region of space
$8,000h^{-1}\mathrm{Mpc}$ (comoving) across; a region large enough to contain the
entire visible Universe older than 2.8 billion years in a light cone to a
redshift of 2.3 for an observer at the center of the simulation volume.
The simulation carried out $3.14 \times 10^{20}$ floating
point operations (0.3 zettaflops).  We saved 16 particle dumps
totaling 540~Tbytes, as well as 69 subsamples of the data totaling 34
Tbytes, and 34 Tbytes of data in two light cones (one from the center, and one from the lower left corner).  Had we attempted the same
calculation with a simple $O(N^2)$ algorithm, it would have taken
about ten million times as many operations and approximately 37 thousand years
on the same hardware to obtain the answer.
During the initial stages of the simulation, a single timestep
required about 110 seconds, for a performance of 5.9 Petaflops.  At
the end of the simulation, where significant clustering increases the
tree traversal overhead we perform a timestep in 135 seconds, for a
performance of 4.3 Petaflops\footnote{We perform somewhat more flops per timestep at late times}.

Our aggregate performance over the entire 33 hour and 50 minute period
was 2.58 Petaflops, due to nearly 40\% overhead from disk I/O.  This
overhead was due in large part to the fact that MPI I/O on the Titan
system was limited to only 160 of the roughly 1000 Lustre
Object Storage Targets (OST) on each of the two Titan Atlas
filesystems.  We have demonstrated I/O performance using equivalent
POSIX I/O benchmarks nearly 8 times higher, which will reduce I/O
overhead to less than 5\% for our next run (with an as yet
undetermined cost in development time to replace what we thought was
the proper MPI interface to use to avoid performance surprise).  We also
saved more checkpoint files than were necessary, given the new input
(surprising to us!) that we encountered no failures during a 34 hour
run.

In addition to the \texttt{ds14\_a} simulation, we have performed a number
of lower particle count simulations with progressively better spatial and
mass resolution. These simulations have thus far been used to provide means
for convergence tests, but in and of themselves enable a number of additional
projects.  At this time, only $z=0$ data is available, but we expect to release
additional snapshots, halo catalogs, and merger trees in the future.

\section{Early Science Results}
\label{sec:DataVerificationValidation}

While this manuscript's primary purpose is to announce the Dark Sky
Simulations campaign and encourage peer review of the data and
development of compatible tools and feedback on its direction, we
present present several state-of-the-art scientific results that have
been enabled by these simulations.  In the following two sections we
present the $M_{200b}$ halo mass function, followed by 
an initial power spectra analysis and a comparison to the Planck
Sunyaev-Zel'dovich all-sky galaxy cluster catalog.

\subsection{Halo Mass Function}
\label{sub:HaloMassFunction}

One of our primary scientific goals is to improve upon the spherical
overdensity (SO) mass function derived in \cite{tinker08}, which has a
quoted accuracy for virial masses of 5\% up to $10^{15} h^{-1}
M_\odot$.  To give some sense of the computational advances enabling
this improvement, in the 30 years since the seminal work of
\cite{press74}, the number of particles in a simulation has increased
from one thousand to one trillion (a factor of $10^9$).  We rely on
progress in creating initial conditions~\citep{crocce06,blas11},
algorithms~\citep{warren13}, computing capability~\citep{bland12} and
analysis~\citep{turk11a,behroozi13} to increase the volume simulated
and analyzed at high fidelity.  We also use a better-constrained input
cosmology~\citep{2011MNRAS.416.3017B,bennett12,2014MNRAS.441...24A,delubac14,bicep2collaboration14,audren12a,warren14a},
reducing the effects of non-universality in applying the
\cite{tinker08} mass function to the current standard cosmological
model.  While the mass function is almost independent of epoch,
cosmological parameters and initial power spectrum (as suggested
by~\cite{press74} and demonstrated by~\cite{jenkins01} at the 10-30\%
level), more detailed work has shown this universality is not obtained
more precisely \citep[e.g.][]{tinker08, courtin11, bhattacharya11}.

\begin{figure*}[htp]
\begin{center}
\dofig{
    \includegraphics[width=1.0\textwidth]{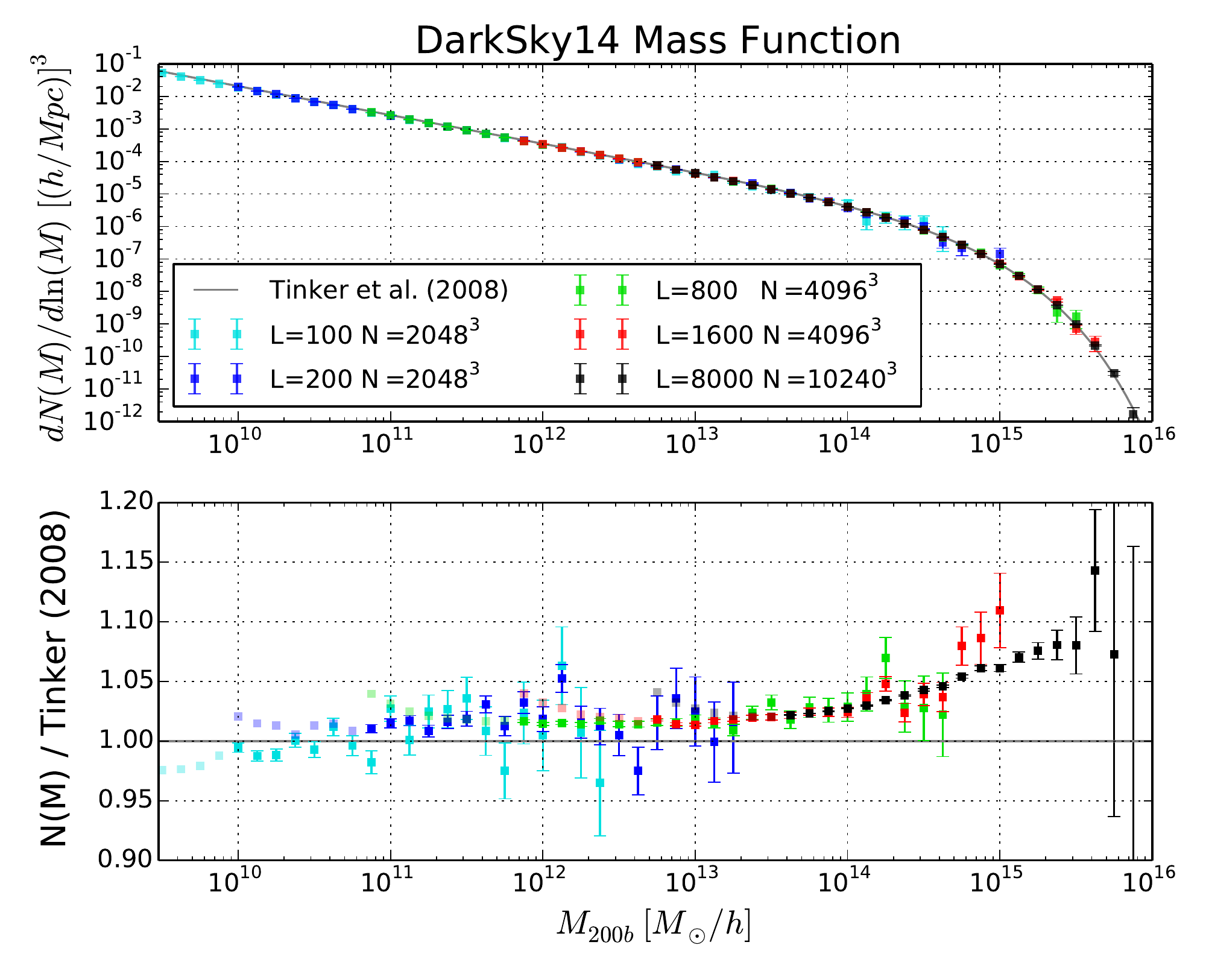}
}
\end{center}
\caption{ The top panel shows the number of halos per mass bin
  compared with \citet{tinker08} for the five Dark Sky Simulations
  introduced in this paper, over six orders of magnitude in halo mass.
  Each decade in mass is split into eight logarithmically spaced
  bins. The error bars indicate the Poisson error for each bin.  The
  ratio between the number of halos and \citet{tinker08} is shown in
  the lower panel, on a linear y-axis.  The simulations are internally
  consistent at the 1\% level for halos with more than $\approx$ 800
  particles.  Lighter points in the lower panel denote halos with
  100-800 particles, which are consistent at the 3\% level.}
\label{fig:hmf}
\end{figure*}

For large masses, the \cite{tinker08} fit obtains most of its
statistical power from the 50 realizations of the larger-volume WMAP1
L1280 ($(1280h^{-1}\mathrm{Mpc})^3$) simulations.  Since the WMAP1
cosmology lies considerably outside current observational constraints,
any non-universality in the translation from WMAP1 to the currently
favored cosmology should be added to the 5\% error budget.  We analyze
here roughly 5 times as much volume (512 $h^{-3}\mathrm{Gpc}^3$) as
\cite{tinker08}, with particle masses ($3.9 \times 10^{10} h^{-1}
M_{\odot}$) about 15x smaller than the L1280 simulations.  It remains
to be demonstrated if we are within reach of the sub-percent
statistical accuracy required for discriminating Dark Energy models
from future surveys, as calculated by~\cite{wu10}.

\cite{reed13} present a number of detailed tests of requirements for
the recovery of the mass function to percent level accuracy, and
present parameter guidelines for doing so.  Their requirements for the
``very challenging prospect'' of a simulation with a light cone to
replicate the volume accessible to future cluster surveys with
sufficient mass resolution to avoid sensitivity to simulation
parameters are essentially met by our current simulation (we obtain
800 particles per halo of mass $M = 10^{13.5}h^{-1}M_\odot$ rather
than the suggested 1000, but our mass resolution is better than those
of the large-volume tests they present, which reduces the sensitivity
somewhat).  As a stringent test of both simulation and halo-finding
consistency, we use the ROCKSTAR halo finder~\citep{behroozi13} to
select halos in a series of five simulations with particle masses
varying between each by a factor of 8, for an overall dynamic range in
mass of 4096 ($9.53 \times 10^6 h^{-1} M_\odot$ to $3.90 \times
10^{10} h^{-1} M_\odot$).

The largest systematic error we have identified primarily affects the
mass of the smallest halos (less than 800 particles), and is related
to the initial evolution of clustering on the inter-particle scale.
Representing a mode with too few particles results in it growing more
slowly than it should.  This loss of small-scale power is made worse
by starting at higher redshift (because the error has longer to
accumulate before non-linear clustering takes over) so the use of 2LPT
initial conditions is required to allow a simulation to start at an
appropriate redshift.  However, even 2LPT initial conditions will lose
small-scale power if started at a redshift higher than they need.  The
correction term to the truncation error is exactly the same
mathematical form as deconvolving a cloud-in-cell density
interpolation, so we enable the \verb-CORRECT_CIC- flag in the initial
conditions generator, even when starting with particles on a grid.
Note that this correction will not apply to codes which compute
short-range forces with a Fourier method, since the force kernel may
already apply ``sharpening'' with the same effect.

In order to compare our theoretical and numerical models with the
universe, we require a reliable calibration of the correlation between
an observable and a quantity measurable from our simulation data.
Since observational data do not provide halo masses neatly derived at
a fixed overdensity, the relationship between measurable and
observable is non-trivial.  At the levels of accuracy required for
next-generation surveys, the details are important.  In particular,
the best definition of ``halo mass'' in a simulation will be
influenced by the computational complexity of the cosmological
parameter estimation it is used for, as well as the ability to connect
it with an observed halo mass.  Even algorithmic choices will
affect the precise definition of a spherical overdensity halo mass at the 1\%
level~\citep{knebe11}.  A measure which yields mass functions which are
self-similar and parameterized simply in terms of the details of the
cosmological parameters and initial conditions would be ideal, but
none of the currently favored measures of halo mass meets this goal.
Part of our motivation for providing raw particle data and the
framework to apply different mass measures is to encourage exploration
of alternative definitions which may suit either theoretical models or
particular observational programs better than the currently used
spherical overdensity.

We have also identified halos in the light cone dataset, which places an
observer at the center of the \texttt{ds14\_a} simulation volume.  At a radius
of $4 h^{-1}~\mathrm{Gpc}$, this yields data out to $z\sim2.3$. Within this
volume, we identify 1.85 billion halos with 20 particles or more.
This data can be used in a variety of contexts, both observational
and theoretical.  Observationally, this dataset can be used for creating mock
galaxy catalogs \citep{2002ApJ...575..587B}, weak lensing predictions, and large
scale clustering predictions. Of particular interest is the distribution
and clustering of the largest halos in the Universe.  Figure \ref{fig:planck} shows
halo masses as a function of redshift in the \texttt{ds14\_a\_lc000} dataset.
The largest halo has an $M_{200b}$ mass of $4.35 \times 10^{15} h^{-1} M_\odot$,
and is at a distance of $631h^{-1}Mpc$ (a redshift of $z=0.22$). This can be compared with theoretical
predictions such as \citet{holz12} directly, where they determined
the most massive object in the universe should be $M_{200b} = 3.8\times10^{15}
M_\odot$.  We note that they assumed a slightly different cosmology, and notably 
a slightly lower $\sigma_8=0.801$. A detailed analysis of the statistics of 
the most massive halos is forthcoming.\\

\subsection{Power Spectra}
The matter power spectrum is a convenient statistic for probing the
time-evolution of spatial clustering of matter in the Universe.
Measuring the matter and galaxy power spectrum (and its inverse Fourier
transform, the correlation function), on linear and non-linear scales,
both directly and indirectly, is one of the major goals of many
ongoing and proposed observational projects; Pan-STARRs~\citep{kaiser02},
, BOSS~\citep{2014MNRAS.441...24A}, DES~\citep{descollaboration05}, 
SPT~\citep{hou14}, WiggleZ~\citep{marin13}, Planck~\citep{planckcollaboration13}, 
LSST~\citep{ivezic08}, SKA~\citep{dewdney09}, DESI~\citep{2013arXiv1308.0847L}, 
Euclid~\citep{amendola13}.  Notable theoretical approaches for calculating
the non-linear power spectrum from an initial linear spectrum and the
cosmological parameters are the `scaling Ansatz' of \cite{hamilton91}
and its extension to the HALOFIT model of \cite{smith03}, with further
recent refinement by \cite{takahashi12}.  At mildly non-linear scales
perturbation theory approaches have been successful
(e.g. \cite{taruya12}).  Other approaches are based on fitting the
results of N-body simulations with an ``emulator'' 
\citep{heitmann08,heitmann14} or neural network~\citep{agarwal14}.

\label{sub:PowerSpectra}
\begin{figure*}[htp]
\resizebox{1.0\textwidth}{!}{\includegraphics{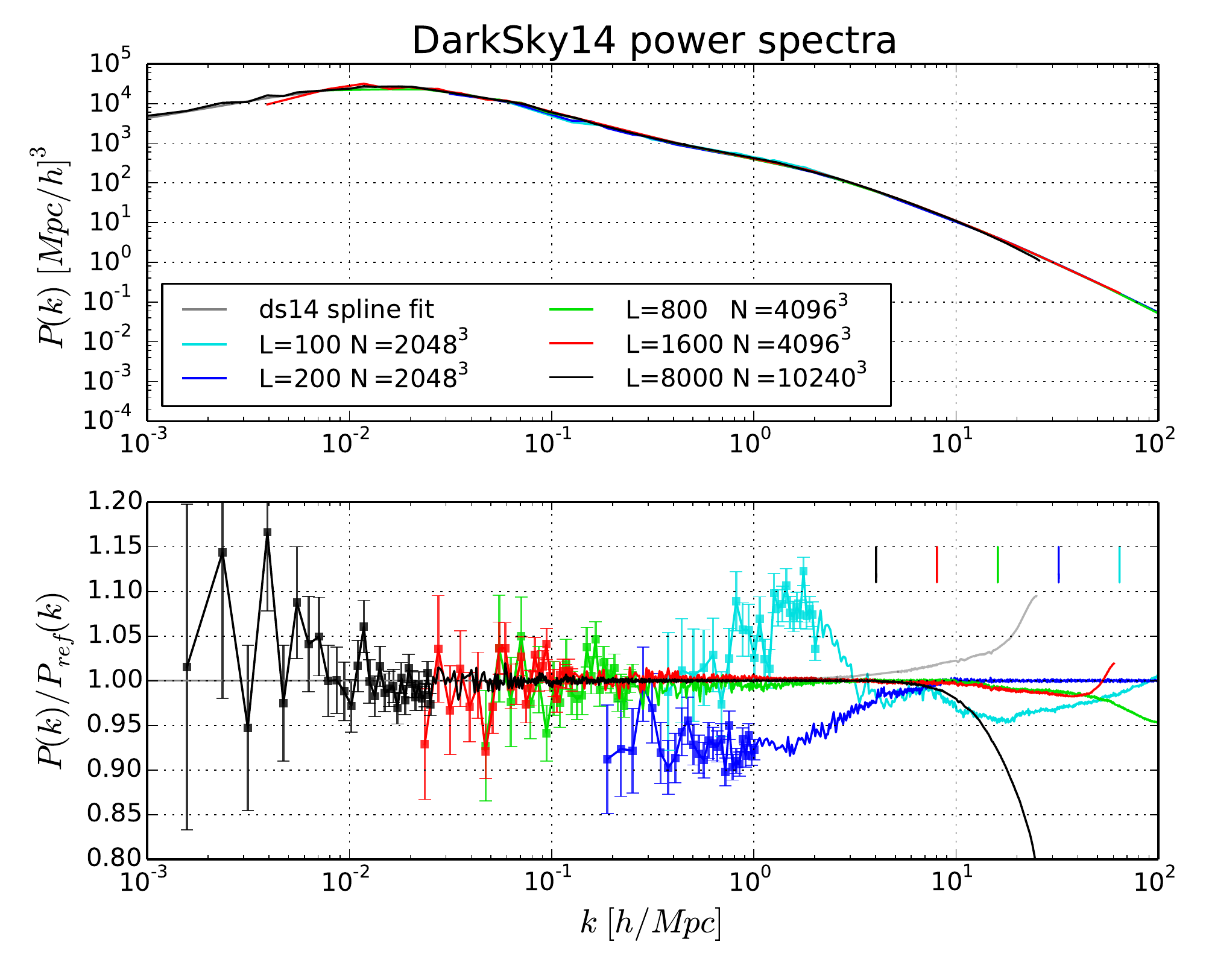}}
\caption{Power spectra at $z=0$ for the five simulations presented here. Top: As a
  function of wavenumber, $k$, the measured power spectra are shown.  Lower: The power
spectra are shown relative to a spline fit of the combined simulation data.  Error bars
on the largest spatial modes show their expected variance.  Vertical ticks
indicate the Nyquist frequency of the initial conditions for each simulation.  Shot
noise corrections are not applied, except for the $8000 h^{-1}\mathrm{Mpc}$ box, where the spectrum is
shown with and without (light gray) a shot noise correction.
}
\label{fig:powspec}
\end{figure*}

\begin{figure*}[htp]
\resizebox{0.48\textwidth}{!}{\includegraphics{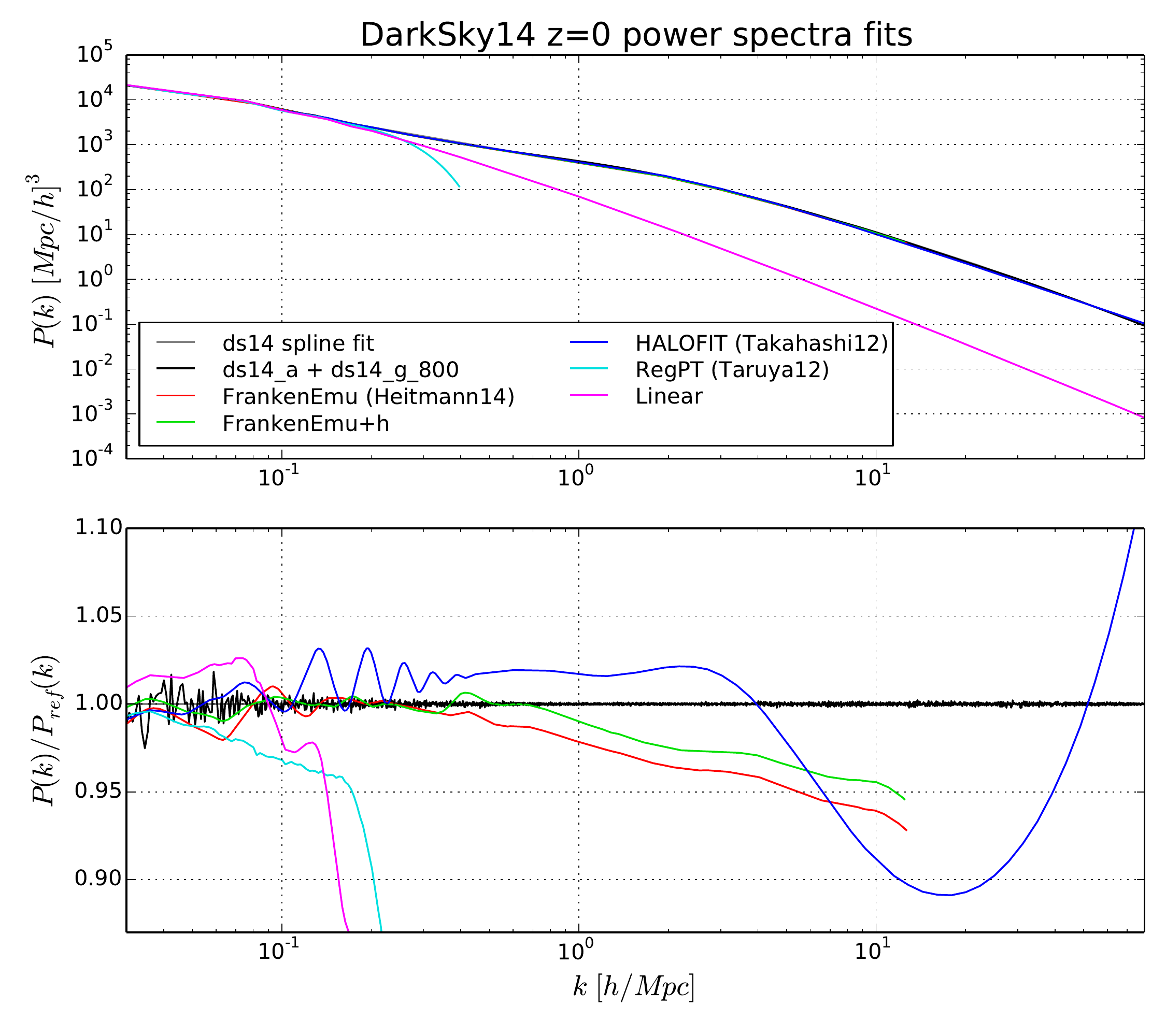}}
\resizebox{0.48\textwidth}{!}{\includegraphics{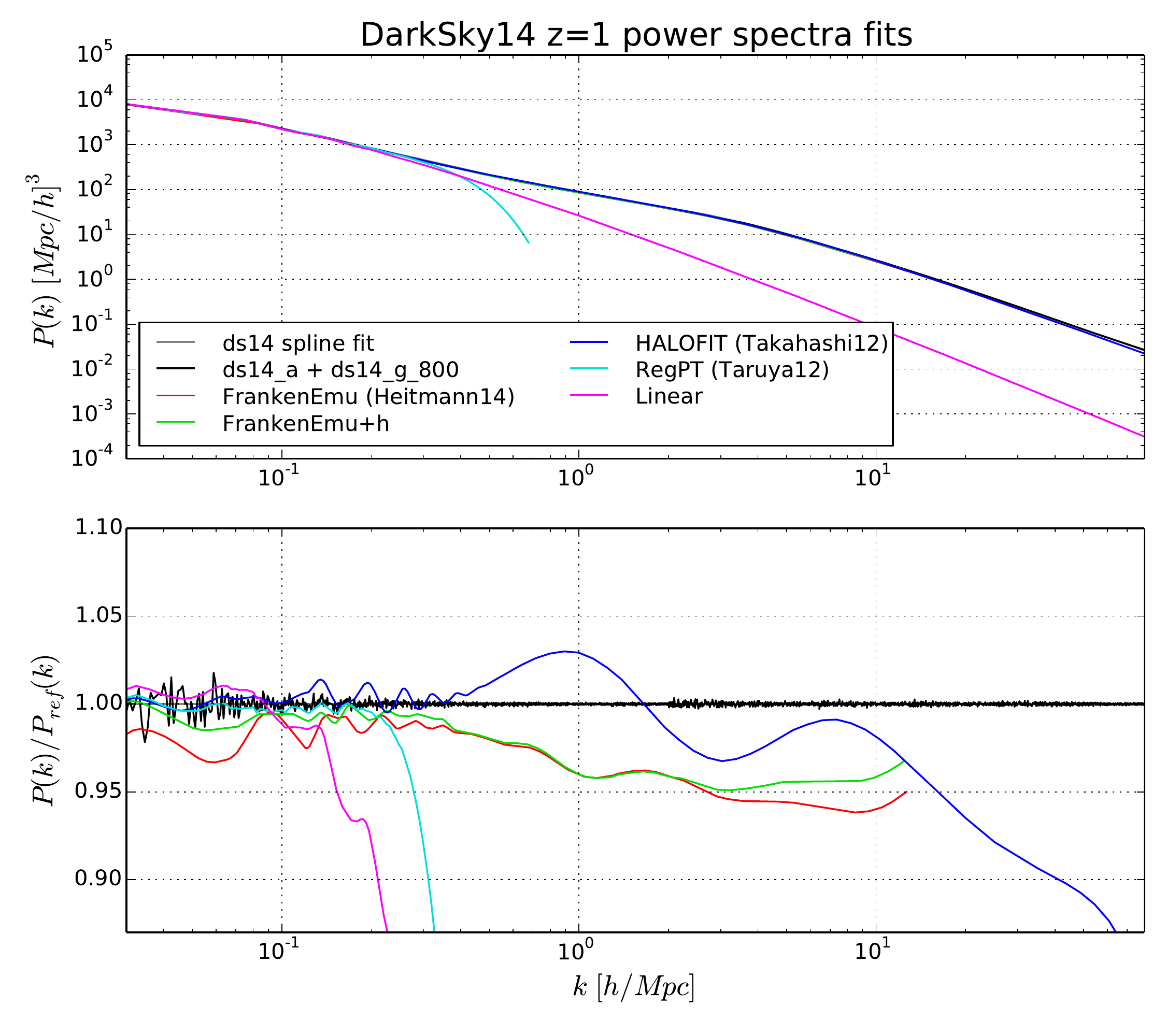}}
\caption{
Top panels: As a function of wavenumber, $k$, the measured power spectra of the
\texttt{ds14\_a} simulation and various fits are shown. 
Bottom panels: The power spectra are shown relative to a spline fit of the combined
simulation data. The left and right panels correspond to $z=0$ and $z=1$ data, respectively. 
The agreement
at large scales with the FrankenEmu+h emulator is very good.
HALOFIT differs from our results at the 1-2\% level, both in
matching the wiggles at the BAO scale and by a small normalization
offset up to $k=3$, and a larger deviation at higher $k$ for the $z=0$
results.  In the right panel of the figure ($z=1$) where perturbation theory would be expected
to match to higher $k$, RegPT matches our results extremely well up to
$k=0.3$.}
\label{fig:powspec3c}
\end{figure*}

In addition to being sensitive to cosmological parameters, the power
spectrum is influenced by the volume, mass resolution and code
parameters used for the simulation.  A necessary (but not sufficient)
condition for a well-behaved simulation model is that the measurement
of relevant physical quantities be independent of these non-physical
parameters.  It is often difficult to measure these sensitivities,
since performing a simulation with a large volume (to reduce
statistical errors) and high mass resolution (to probe small spatial
scales) soon becomes prohibitively expensive to compute.

Within the current suite of Dark Sky Simulations, we have both
large volume and good mass resolution, providing an opportunity
to check the sensitivity of the power spectrum to non-physical
parameters.  These tests are complementary to the previous comparison
of the power spectrum evolved from the same initial conditions using
the \2HOT and GADGET~\citep{springel05} codes that were presented in
\cite{warren13}.  In Figure~\ref{fig:powspec} we show the power
spectrum measured from five simulations differing in mass resolution
by factors of 8.  It is to be noted on the log-log scale of the top panel that
the measured power spectra differ by less than the width of the line
over most of the spatial scale.  To minimize distortions from the
precise form of the mass interpolation at small scales, we perform
large ($4096^3$--$8192^3$) FFTs, and perform multiple FFTs with the spectrum
folded up to a factor of 8 to probe high $k$ without approaching the
Nyquist limit of the FFT.

The lower panel of Figure~\ref{fig:powspec} shows the power spectra on
a linear scale, divided by a spline fit to our data.  The spline uses
the $8000h^{-1}\mathrm{Mpc}$ data up to $k=2$, $800h^{-1}\mathrm{Mpc}$
from $k=2$ to $k=10$, and $200h^{-1}\mathrm{Mpc}$ data beyond
$k=10$. The most obvious 10\% difference between simulations is due to
the ``cosmic variance'' of the smaller 100 and $200h^{-1}\mathrm{Mpc}$
volumes.  The three larger volume simulations show remarkable
agreement (well below the 1\% level) over the range from where the
mode variance is small, down to spatial scales near the mean
inter-particle spacing. The vertical ticks represent the Nyquist
frequency of the initial conditions, representing the spatial scale
below which there is no information when the simulation begins.  Our
convergence results near the mean inter-particle spacing echo the
conclusions of~\cite{splinter98}, that increased force resolution must
be accompanied by sufficient mass resolution, and that results below
the inter-particle scale are subject to larger systematic errors.

In Figure~\ref{sub:PowerSpectra} we show our results compared with
FrankenEmu~\citep{heitmann14}, HALOFIT~\citep{takahashi12} as
generated from the CAMB code~\citep{lewis00}, RegPT~\citep{taruya12}
and the linear power spectrum.  The matching of the wiggles at the BAO
scale in HALOFIT differs from our results at the 2\% level.  The
agreement at large scales with the FrankenEmu+h emulator is very good.
FrankenEmu also matches up to k=10 within their quoted 5\% accuracy.
Since FrankenEmu was not calibrated using this specific cosmology
within the emulator framework, errors at the 5\% level are expected by
$k=10$, as was the case for their \texttt{M000} test cosmology.
Perturbation theory results from RegPT are not a particularly good
match at $z=0$ where the system is highly evolved, but at $z=1$ in the
right panel of the figure where perturbation theory would be expected
to match to higher k, RegPT matches our results extremely well up to
$k=0.3$.

\subsection{Comparison to Planck SZ Cluster Catalogs}

One motivation for the large $8^{-1}\mathrm{Gpc}$ volume is given by full sky
galaxy cluster surveys.  For example, \citet{reed13} point out that the
full-sky volume corresponding to $z=2$ could be achieved through a 
simulation with a box size of $8h^{-1}\mathrm{Gpc}$ on a side, with enough mass
resolution to populate halos with $M \gtrsim 10^{13.5} h^{-1} M_\odot$ with 
$N\sim1000$ dark matter particles.  \texttt{ds14\_a} nearly achieves this, having
$N\sim800$ particles in the desired mass halo.  As such, this simulation provides
a unique opportunity for comparison with on-going surveys. In order to showcase 
this opportunity, we demonstrate its use in creating a simple mock SZ cluster catalog.
Properly constructing a full SZ mock catalog is beyond the scope of this work, and
the current presentation should be viewed as a demonstration.

\begin{figure*}[htp]
\centering
\resizebox{0.8\textwidth}{!}{\includegraphics{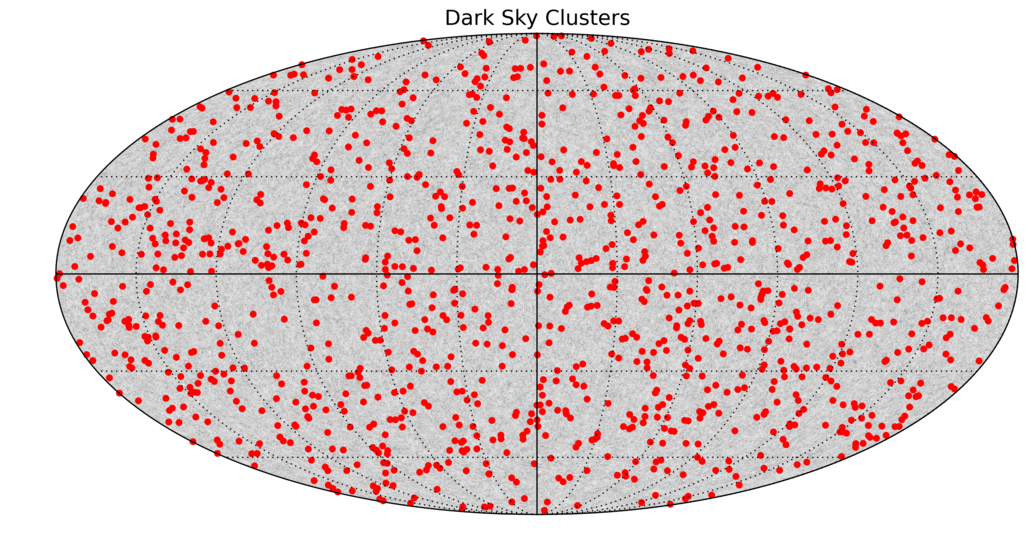}}
\resizebox{0.8\textwidth}{!}{\includegraphics{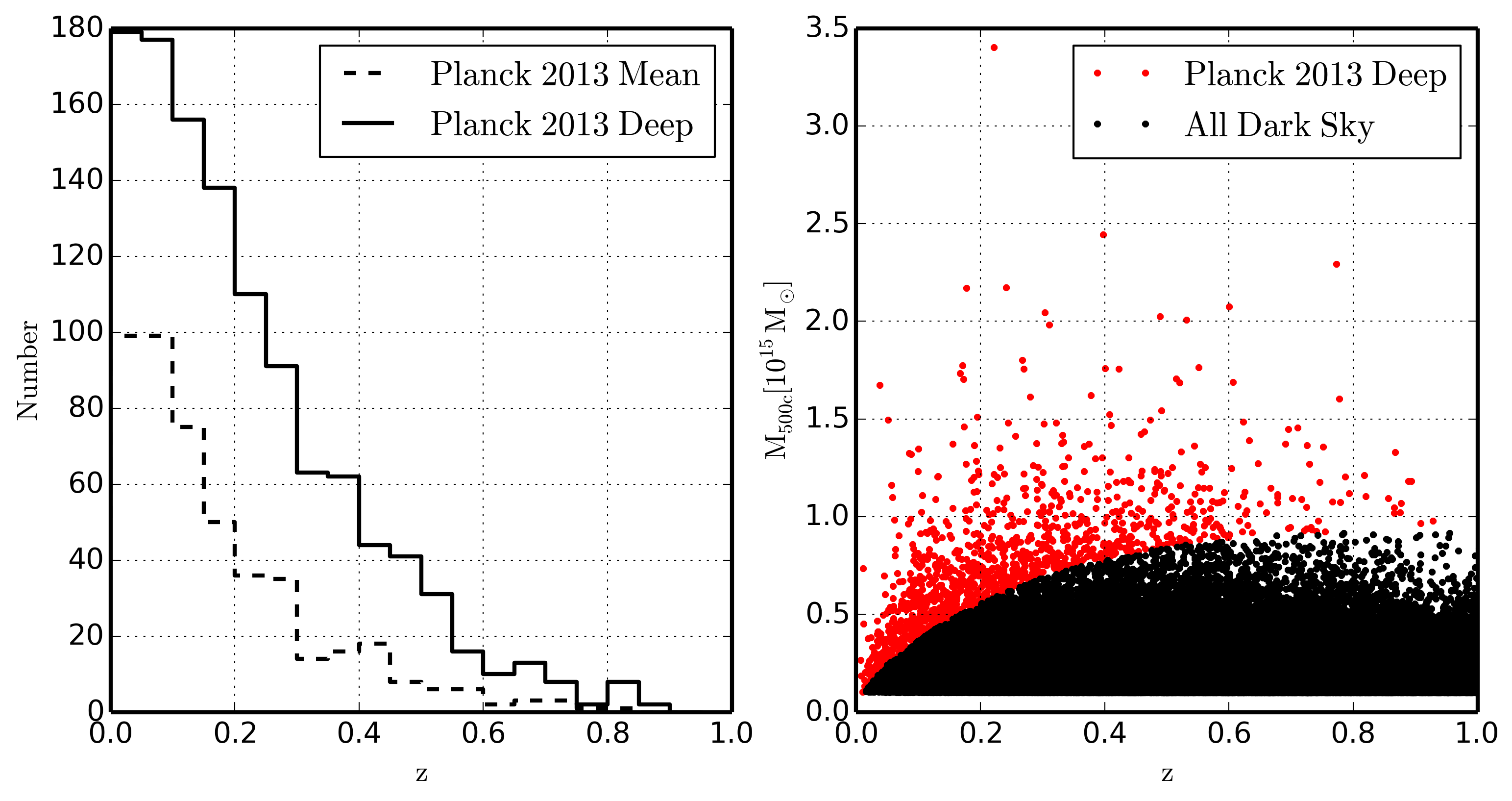}}
\caption{
  Top: Positions of all simulated galaxy clusters that would be visible in an
  all-sky survey of the \texttt{ds14\_a} light cone dataset using the Planck
  $2013$ Deep estimates.  Bottom-left: The number of galaxy clusters visible as
  a function of redshift (in $\Delta z=0.05$ bins), assuming limits from two
  minimum mass estimates from \citet{2013arXiv1303.5080P}.  Bottom-right:
  Masses of galaxy clusters in the light cone dataset as a function of
  redshift. Red points indicate those that would be visible with the Planck
  Deep filter, while black denote all Dark Sky clusters with $M_{500c} >=
  10^{14} M_\odot$.
}
\label{fig:planck}
\end{figure*}

We use the results shown in Figure 3 of \citet{2013arXiv1303.5080P} to construct
a minimum mass filter (as defined w.r.t $M_{500c}$) as a function of redshift. Here
we use both the ``Deep'' and ``Mean'' estimates from their study. In Figure \ref{fig:planck},
we show the positions on the simulated sky, the histogram of clusters as a function
of redshift, and the masses of the galaxy clusters as a function of redshift. 
Figure \ref{fig:planck} can be directly compared with Figure $2$ and $24$ from
\citet{2013arXiv1303.5089P}. We see a quite striking agreement between the Planck ``Mean''
estimate with their (Figure $24$) Planck-MCXC sample, and between the Planck ``Deep'' estimate
with the combined set of their ``Planck clusters with redshift'' distribution.

\section{Available Data Products}
\label{sec:DataProducts}

In this first data release our goal is to provide enough public data
to enable state-of-the-art scientific research, as well as provide a testbed
for the public interface.  A list of available data products as of July 2014 is 
listed in Table~\ref{tab:datasets}.  Raw data can be downloaded directly from
the web without authentication. While the exact HTTP address may migrate as
server usage is evaluated, we will keep an updated python package
\texttt{darksky\_catalog}\footnote{\url{https://bitbucket.org/darkskysims/darksky\_catalog}}
that can be used to
alias a given simulation data product to its current uniform resource locator
(URL). Updated information will also be kept on our project website
(\url{http://darksky.slac.stanford.edu}). As well as directly
accessing data through a web browser, we provide advanced access methods
through \texttt{yt}, detailed at the end of this Section.

\begin{table}[htb]
\centering
\begin{tabular}{|c|c|r|}
\hline  Simulation & Description & Size \\ \hline
\hline  
ds14\_a & $z=0$ Particle Data         & 34 TB \\ \hline
ds14\_a & $z=0$ Halo Catalog          & 349 GB \\ \hline
ds14\_a & Lightcone Data $(z < 2.3)$  & 16 TB \\ \hline
ds14\_a & Lightcone Halo Catalog      & 155 GB \\ \hline
ds14\_g\_1600\_4096 & $z=0$ Particle Data  & 2 TB \\ \hline
ds14\_g\_800\_4096  & $z=0$ Particle Data  & 2 TB \\ \hline
ds14\_g\_200\_2048  & $z=0$ Particle Data  & 256 GB \\ \hline
ds14\_g\_100\_2048  & $z=0$ Particle Data  & 256 GB \\ \hline
\end{tabular}\\[.25in]
\caption{Available datasets as of July, 2014. A full listing is available at the
  project website.}
\label{tab:datasets}
\end{table}

\begin{figure}[htp]
\begin{center}
\dofig{
    \includegraphics[width=0.5\textwidth]{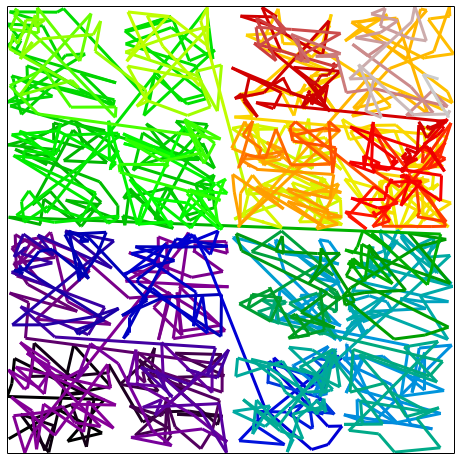}
}
\end{center}
\caption{Morton ordered data. A small subset of particles from the \texttt{ds14\_a} simulation,
connected in order as they are sorted on-disk. The color indicates increasing offset in the file.
The recursive ``Z-order'' pattern can be seen. Spatial queries that are aligned with octants result
in a single read, after which particles may be filtered quickly in-memory.}
\label{fig:morton_particles}
\end{figure}

The datasets in this EDR fall roughly into three categories: raw particle data,
halo catalogs, and reduced data. Raw particle data, and halo catalogs are stored
in single \texttt{SDF} files for each snapshot/redshift. This leads to individual
files that are up to 34 TB in size. In addition to the primary \texttt{SDF}
file, we create a ``Morton index file'' (\texttt{midx}), detailed in~\ref{sub:SDF} that allows
for efficient spatial queries (see Figure~\ref{fig:morton_particles}).  By combining the raw particle
dataset and a \texttt{midx} file, sub-selection and bounding box queries can
be executed on laptop-scale computational resources while addressing an arbitrarily large simulation.
Particle datasets store particle position
$(x,y,z)$ and particle velocity $(vx, vy, vz)$ each as a 32-bit \verb~float~,
and a particle unique identifier, $id$, as a 64-bit \verb~int~. Halo catalogs store
a variety of halo quantities calculated using Rockstar~\citep{behroozi13},
described in Section \ref{sub:HaloFinding}.

In addition to raw particle snapshots, during runtime we output two light cone
datasets, one from a corner of the simulation volume and one from the center. At
this time we are releasing the lightcone from the center of the box, which covers
the full sky out to a redshift of $z=2.3$.  Halos found from this lightcone are
also released.  A slice through the halos found in the lightcone dataset is shown
in Figure \ref{fig:lightcone}. 

\begin{figure*}[htp]
\begin{center}
\dofig{
    \includegraphics[height=0.9\textheight]{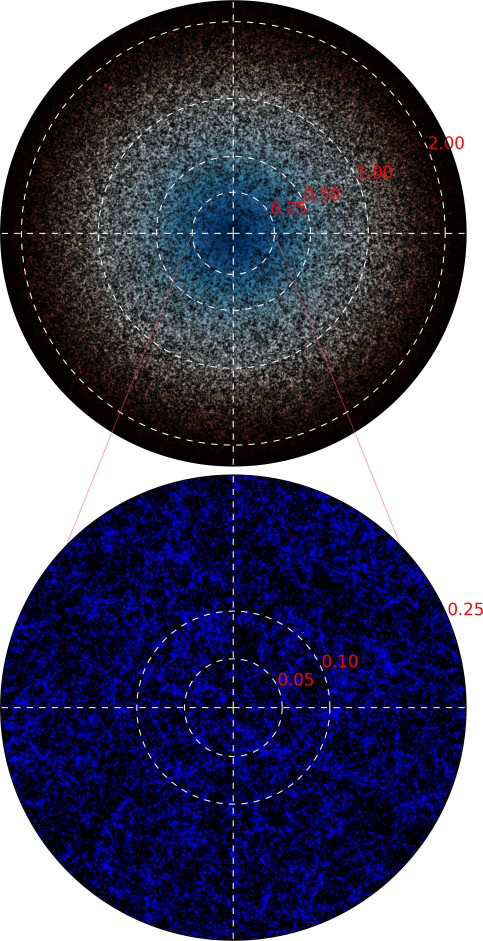}
}
\end{center}
\caption{A $10\deg$ slice of halos above $10^{13} M_\odot$ found in the \texttt{ds14\_a} 
  lightcone dataset. Each point is colored by its proper radial velocity from the center.
  The top panel shows the full radial extent of the halo catalog, which covers $z<2.3$.
  The bottom panel is a zoom on $z<0.25$. This represents $1/18th$ of the halos in the
  lightcone dataset.
}
\label{fig:lightcone}
\end{figure*}

\begin{figure*}[htp]
\begin{center}
\dofig{
  \includegraphics[width=0.8\textwidth]{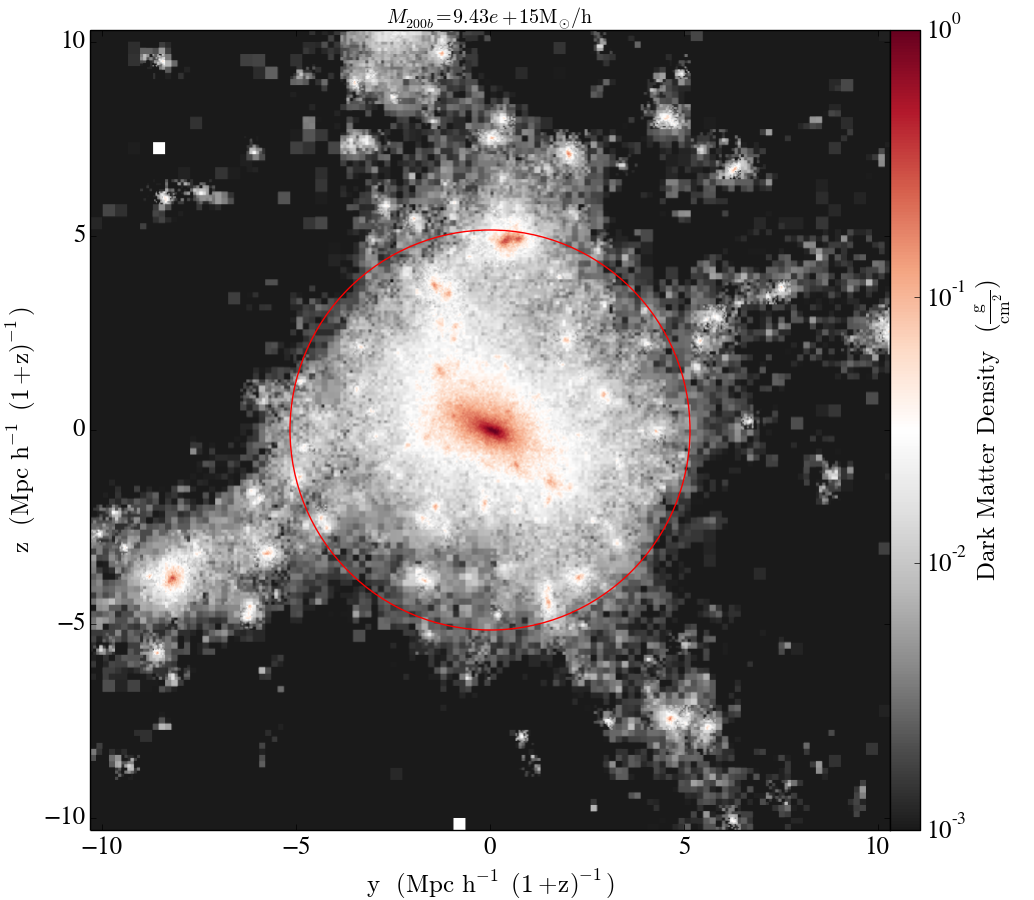}
}
\end{center}

\caption{Halo loaded remotely by \yt~that uses the the darksky\_catalog 
abstraction layer to redirect to datasets hosted on the WWW. Code used
to generate the figure can be found in the project repository at
\url{http://bitbucket.org/darkskysims}.}
\label{fig:halo_particle_load}
\end{figure*}

Data can be browsed and directly downloaded using a common web browser.
However, given the data volume of the raw particle data, we do not recommend
directly downloading an entire file. Instead, we have provided a simple remote
interface through the commonly used \texttt{yt} analysis and visualization software.
In this way, a researcher may directly query a sub-volume of the domain and download
only what is needed for their analysis. These fast spatial queries are enabled
by storing the majority of the data products (raw particle, halo catalogs) in
``Morton order'' on the server.  Each particle is described by 32 bytes, and therefore
loading a million particles around a halo in \texttt{ds14\_a} requires downloading $32 \rm{MB}$. The
average broadband internet speed in the U.S. (July, 2014) is 25.1 Mbps\footnote{\url{http://www.netindex.com/download/2,1/United-States/}}, meaning that an average household could
load this data in less than 10 seconds. Well-connected institutions would presumably
be faster. Given
a bounding box and Morton index file, offsets and lengths can be used with an
HTTP range request to return the desired data.  These queries are cached
locally such that once loaded, particle data can be analyzed in-memory without
any special requirements on the server.  An example analysis is shown in Figure
\ref{fig:halo_particle_load}, that utilizes helper functions from the
\texttt{darksky\_catalog} to remotely load the particles around the most
massive halo in the \texttt{ds14\_a} simulation into \texttt{yt}, and project
the dark matter density along the line of sight.

\section{Community Standards}
\label{sec:CommunityStandards}
The full stack of open-source software utilized to create this
data---from the operating system, to the file systems, the compilers,
numerical libraries, and even the software used to typeset this
paper---is the product of the efforts of thousands upon thousands of
individuals.  In most of these cases, we are afforded the benefit of
decades of an enormous and largely intangible investment in shared
software development.  We are releasing this data in the spirit of
open science and open software development, with the intention of
making it \textit{immediately} usable to individuals with a wide range
of skills and interests, encouraging the sharing and reuse of derived
data products, and perhaps most importantly, removing ourselves as
obstacles to its use.  By doing so, we are attempting to participate
in a material way in the development of future scientific endeavors,
as we have benefited from the open release of software and data in the
past.  The value and quantity of this data vastly exceeds our ability
to mine it for insight; we do not wish to see its utility bottlenecked
by our own limitations.

On one hand, we would like the data to be usable with as little
bureaucratic or viral licensing overhead as possible. On the other, we do not
wish that it be used for unfair advantage.  To support these goals, we
hope to foster the continued growth of a community of individuals
using the data, contributing to exploration of the data, and
developing tools to further analyze and visualize this and other data
within the field.  We suggest below some guidelines for usage to
foster this community, and as a basis for further discussion of this
new facet of data-intensive scientific inquiry enabled by the incredible
progress in computational hardware and software.  We believe that the longevity and
success of this and future endeavors in the release of data for open
science will be influenced by the spirit of openness and collaboration
with which this and other path-finding projects are received.

\textbf{Citation:} Key to ensuring both ``credit'' (see \citet{katz14}
for more detailed discussion) and good feelings between individuals is
proper attribution of other, related or foundational work upon which
discovery is based.  We expect that discoveries and data products that
are enabled though data from the Dark Sky Early Data Release cite this
manuscript.  This will enable others to track the provenance of the
data products, replicate the discovery themselves, and also ensure
that any impact this work has can be measured and evaluated.  The
canonical references to work related to this release are
\2HOT~\citep{warren13}, \yt~\citep{turk11a} and SDF \citep{sdf2014}.
Work using halo catalogs should include~\citep{behroozi13}.  In
addition, we appreciate being informed if we have overlooked the
citation of a significant contributor to our own work.  Particular
care is required for highly interdisciplinary scientific and
computational research, since standards for citation do vary across
fields.

\textbf{Collaboration:} In some instances, enabling the use of this
data to address new questions will require effort that is best
formalized through the establishment of a scientific collaboration,
with the intent being co-authorship for those who provide significant
additional investment.  We welcome additional collaborators, and
can be contacted at the e-mail addresses provided.

\textbf{Mirroring Data:} We encourage other projects and individuals
to share their own copies of the data, and request that they both
notify us (so that we can provide links to the mirrors) and refer back
to the original website (to attribute as well as help ensure that
individuals relying on the mirror can identify ``upstream'' changes or
enhancements).  All data released here is available for both
commercial and non-commercial use, under the terms of the Creative
Commons Attribution 4.0 License;
\url{http://creativecommons.org/licenses/by/4.0/}.

\textbf{Derived Catalogs:} Derived data, such as catalogs, generated from the
datasets released here are not required to be made publicly available, although we
would encourage individuals creating those data catalogs to do so.  Any
individuals generating derived data products from this data may request that
their derived data products either be hosted alongside or linked to from the
canonical data repository.

\textbf{Feature Enhancements:} The source code for each component of the
analysis pipeline of DarkSky is hosted in a publicly accessible,
version-controlled repository.  We welcome and encourage the development of new
features and enhancements, which can be contributed using the pull request
mechanism.  We also encourage feature enhancements be documented and published
(even if through a service such as Zenodo or Figshare) so that contributors
are able to collect credit for their work. The value of this data can be greatly
expanded by enabling the development of platforms on which it can be analyzed
\citep{Turk2014} and through the review and acceptance of feature enhancements
we intend to develop this value.

\textbf{Discussion Forum:} We have established a discussion forum at
darkskysimulations@googlegroups.com and a Google plus community
at \url{http://DarkSkySimulations.info}.  We encourage interested
individuals to subscribe and to post questions, feedback, and general
discussion.

\section{Conclusions}

In this work we present the first public data release from the ongoing ``Dark
Sky Simulations'' cosmological simulation campaign. These first five simulations include
one of the largest simulations carried out to date, with $10240^3$ particles
in an $(8h^{-1}\mathrm{Gpc})^3$ volume.  
The main findings from our work can be summarized as the following:
\begin{itemize}
  \item A single-method hierarchical tree approach to the N-body gravitational problem
    is computationally feasible, accurate, and performant on modern HPC architectures.
  \item Using this method, we've carried out a suite of state-of-the-art cosmological
    simulations, hereafter referred to as the Dark Sky Simulations. 
  \item We present results comparing the mass function and power spectra
    to demonstrate the quality of our simulations.  We find internal
    consistency between different box sizes at the 1\% level over more
    than 3 orders of magnitude in particle number.  Comparisons with
    results in the literature agree at the 1-10\% level depending on scale.
  \item The \texttt{ds14\_a} light cone dataset and associated halo catalog
    provide a unique resource to make predictions for the large
    volumes probed by current and upcoming sky surveys; we
    use all-sky Sunyaev-Zel'dovich effect cluster counts as an example application.
  \item Interacting with data from Petascale supercomputing simulations is
    algorithmically challenging; We have designed and implemented a novel data
    access approach that is simple, extensible, and demonstrated its capability
    of interacting with individual files that are 34 TB in size over the Internet.
  \item We have reduced the time to the dissemination phase of our research by
    providing open access to a significant portion of the raw data from our
    simulations less than three months after the simulation was run, totalling
    more than 55 TB of publicly accessible datasets.
\end{itemize}


We encourage the use and further analysis of these data as well as feedback on
the accuracy and accessibility of the data.  We will be updating and adding to
the contents of the data release, including new simulations, and
encourage feedback on which data products and simulation suites would be most
useful.  We will notify the community through the project website and
discussion forum.  The computational resources required to generate these data
were only a small fraction of our currently allocated computing time, so we
expect the amount of and variety of scientifically useful data products
to grow rapidly.

\section*{Acknowledgments}
We are grateful to Andrey Kravtsov and Matt Becker for
useful discussions regarding the mass function, and particularly thank
Matt Becker for providing data to compare with our early mass function
results.  We thank the Institutional Computing Program at LANL for
providing the computing resources used for the initial development and
simulations for this project.  This research used resources of the Oak
Ridge Leadership Computing Facility at Oak Ridge National Laboratory,
which is supported by the Office of Science of the Department of
Energy under Contract DE-AC05-00OR22725.  We thank Wayne Joubert, our
INCITE project computational project liason at ORNL, and the rest of
the OLCF support staff for their help in utilizing the Titan system.
We thank the scientific computing team at SLAC for their continued
support related to hosting data through the \texttt{darksky} server,
in particular Stuart Marshall, Adeyemi Adesanya, and Lance Nakata.
This research also used resources of the National Energy Research
Scientific Computing Center, which is supported by the Office of
Science of the U.S. Department of Energy under Contract
No. DE-AC02-05CH11231. This research was performed under the auspices
of the National Nuclear Security Administration of the
U.~S.~Department of Energy under Contract DE-AC52-06NA25396.  SWS was
supported by a Kavli Fellowship at Stanford.  MSW gratefully
acknowledges the support of the U.S. Department of Energy through the
LANL/LDRD Program and the Office of High Energy Physics.  MJT was
supported by NSF ACI-1339624 and NSF OCI-1048505. RHW received support
from the U.S. Department of Energy under contract number
DE-AC02-76SF00515, including support from a SLAC LDRD grant, from the
National Science Foundation under NSF-AST-1211838, and from
HST-AR-12650.01, provided by NASA through a grant from the Space
Telescope Science Institute, which is operated by the Association of
Universities for Research in Astronomy, Incorporated, under NASA
contract NAS5-26555.  DEH acknowledges support from the National
Science Foundation CAREER grant PHY-1151836. He was also supported in
part by the Kavli Institute for Cosmological Physics at the University
of Chicago through NSF grant PHY-1125897 and an endowment from the
Kavli Foundation and its founder Fred Kavli.

\bibliographystyle{myabbrvnat} \bibliography{zotero,sws,mjt}

\appendix{
\section{A Brief Tour of the Dark Sky Simulations Early Data Release}

This material is intended to be useful to a wide audience, so it starts
at a very basic level. It ends at a very advanced level. Skip ahead if
you need to, or slow down and review the supplemental material.

\subsection{Get the Release Metadata}

First, you need to get the package with the basic data release. You can
use your web browser download feature to get a .zip or .tar.gz file.
They can be found at
\url{https://bitbucket.org/darkskysims/data_release/downloads} Or, if
you prefer the command line,

\begin{verbatim}
wget https://bitbucket.org/darkskysims/data_release/get/default.tar.gz
\end{verbatim}
The file you retrieve will be named something like
darkskysims-data\_release-dba678c57371.tar.gz and is currently about 4
Mbytes in size. You should unpack it.

\begin{verbatim}
tar xzf darkskysims-data_release-dba678c57371.tar.gz
cd darkskysims-data_release-dba678c57371/ds14_a
\end{verbatim}
If you are familiar with the mercurial version control system, and it is
installed on your system, you can replace the steps above with

\begin{verbatim}
hg clone https://bitbucket.org/darkskysims/data_release
cd data_release/ds14_a
\end{verbatim}
Using mercurial has additional advantages, since it can tell you if any
of the files have been changed or corrupted. Read more about mercurial
at \url{http://mercurial.selenic.com/}

\subsection{Data Exploration}

Let's look at some metadata,

\begin{verbatim}
head ds14_a_1.0000.head
\end{verbatim}
should produce the following output on a Unix-like operating system If
you don't have UNIX, use whatever tool is available to view an ASCII
file.

\begin{verbatim}
# SDF 1.0
int header_len =  2528;
parameter byteorder = 0x78563412;
int version = 2;
int version_2HOT = 2;
int units_2HOT = 2;
int64_t npart = 1073741824000;
float particle_mass = 5.6749434;
int iter = 543;
int do_periodic = 1;
\end{verbatim}
This is an SDF header. Our raw data is distributed in the SDF format.
The data\_release directory just contains the initial part (.head) of
this large file. Otherwise, you would need to wait to download the 34
Terabytes in the ds14\_a\_1.0000 file You can read more about SDF at
\url{http://bitbucket.org/JohnSalmon/sdf}. If you browse further down in
the file (perhaps using the more or less command) you will see,

\begin{verbatim}
double Omega0_m = 0.295037918703847;
double Omega0_lambda = 0.7048737821671822;
double Omega0_b = 0.04676431995034128;
double h_100 = 0.6880620000000001;
char length_unit[] = "kpc";
char mass_unit[] = "1e10 Msun";
char time_unit[] = "Gyr";
char velocity_unit[] = "kpc/Gyr";
char compiled_version_nln[] = "2HOT_nln-1.1.0-17-gbb2d669";
char compiled_date_nln[] = "Apr 19 2014";
char compiled_time_nln[] = "06:34:59";
\end{verbatim}
That is the metadata in the SDF file which describes the cosmological
parameters used, the physical units, and information describing the
exact version of the code. A bit further down is the first structure
declaration,

\begin{verbatim}
struct {
    unsigned int sha1_len;
    unsigned char sha1[20];
}[65536];
\end{verbatim}
This structure describes the layout in the data file of the sha1
checksums. There are 65536 of them, each with the length of data that
was checksummed and the sha1 checksum value. This allows one to verify
the integrity of each segment of the 34 Terabyte file independently.
Since the sha1 values are contained in the .head file in the
data\_release, their integrity can be verified with the checksum
contained within the mercurial repository. A change to any one of the
272 trillion bits in the data file can thereby be detected.

The main purpose of the SDF header is found in the last structure
definition,

\begin{verbatim}
struct {
    float x, y, z;              /* position of body */
    float vx, vy, vz;           /* velocity of body */
    int64_t ident;              /* unique identifier */
}[1073741824000];
\end{verbatim}
This tells us how to read the positions, velocities and identity of the
particles. There are 1073741824000 of them. Where are they? We tell you
in the .url file.

\begin{verbatim}
cat ds14_a_1.0000.url
http://darksky.slac.stanford.edu/simulations/ds14_a/ds14_a_1.0000
\end{verbatim}
The full 34 TB data file is on a machine at Stanford University (feel
free to explore the server at
\url{http://darksky.slac.stanford.edu/simulations/}). You could download it
like any of the millions of other files on the Internet, but we have
better ways.

\subsection{SDF-enabled Exploration}

If you like C and the command line, the SDF library is a good choice.
Check out the development version from our repository and compile it. It
will help for the next step if you have libcurl installed first.

\begin{verbatim}
cd ..
git clone http://bitbucket.org/darkskysims/sdf.git SDF
cd SDF
make
\end{verbatim}
This will build an executable called SDFcvt. It can be used to browse
SDF files. cp SDFcvt /usr/local/bin or elsewhere in your path if you
like. Change back to the ds14\_a subdirectory of the data\_release, and
try it out.

\begin{verbatim}
cd ../ds14_a
../SDF/SDFcvt ds14_a_1.0000.head Omega0_m Omega0_lambda Omega0_b h_100
0.29503791870384699 0.70487378216718222 0.046764319950341277 0.68806200000000006
\end{verbatim}
SDFcvt has parsed the cosmological values you specified from the header.
Now try this,

\begin{verbatim}
../SDF/SDFcvt -n 1 -s 1000000000000 http://dsdata.org/ds14_a_1.0000 x y z vx vy vz
3222417 3812104 2801807.5 -216.09906 -215.322311 273.964539
\end{verbatim}
That is the position and velocity of the trillionth particle in the data
file. -n 1 specifies that you want to read 1 element from the structure
-s specifies the offset in the array (Note that dsdata.org just
redirects to the Stanford web site listed above. Its purpose is to make
URLs shorter, so DarkSky names are easy to type, or fit on a line.)

If you see a message like this, you do not have libcurl installed

\begin{verbatim}
SFhdrio: MPMY_Fread returns -1, errno=22
\end{verbatim}
If you don't have libcurl, read on to see how you can use our python
library to access the data files over the Internet. You can also
download any of the smaller data files and use SDFcvt locally.

\subsection{Python-based Exploration}

If you like Python, we have implemented several methods to interact with
the data, both directly using Numpy arrays and through yt. Let's start
with relatively simple methods of interacting with the data. To start,
you'll need to install a few packages. The simplest method is to use pip
to install some of the basic packages. To install yt, we suggest you
follow their documentation
(http://yt-project.org/docs/dev-3.0/installing.html).

First let's get ThingKing:

\begin{verbatim}
pip install thingking
\end{verbatim}
or install from souce, located at
http://bitbucket.org/zeropy/thingking.\\ThingKing exposes data on the
WWW to Python in a memory-mapped interface.

If you start up a Python interpreter, try the following:

\begin{verbatim}
import thingking
ds14_a = thingking.HTTPArray("http://darksky.slac.stanford.edu/simulations/ds14_a/ds14_a_1.0000")
\end{verbatim}
\texttt{ds14\_a} is now array-like, and you can do things like examine
its size:

\begin{verbatim}
print ds14_a.size

34359739943392
\end{verbatim}
That's an array with 34 trillion elements! Let's look at the first few.
Since we did not prescribe a type to the HTTPArray, it defaults to being
an array of characters. That means that if you try to print, say, the
first 10 elements, you'll see each character:

\begin{verbatim}
print ds14_a[:10]

[('#',) (' ',) ('S',) ('D',) ('F',) (' ',) ('1',) ('.',) ('0',) ('\n',)]
\end{verbatim}
That is not so useful, but if we print the data attribute that hangs off
\texttt{ds14\_a}, we get something more readable:

\begin{verbatim}
print ds14_a[:10].data

# SDF 1.0
\end{verbatim}
Congratulations, you can now examine any part of a 34 TB file that you'd
like. If you were inclined, you could use the information in the
header, like the \texttt{header\_len} value and the sizes of the
variables to access the particle data. However, we've implemented all of
that within yt, and suggest we move on to there to access the particle
data in Python.

\subsection{DarkSky \& yt}

We are working towards integrating our extensions to yt into the main yt
development repository. Until then, we are maintaining a separate fork,
which you can get by pulling changes from the repository at
\url{http://bitbucket.org/darkskysims/yt-dark} into your local yt repository.
You may also wish to just download a separate clone of this repository.

\begin{verbatim}
hg clone http://bitbucket.org/darkskysims/yt-dark
cd yt-dark
python setup.py install  # or python setup.py develop 
\end{verbatim}
At this point, it would also be most useful to download the example
scripts at \url{http://bitbucket.org/darkskysims/darksky_tour}. Like most
things, you can do this through downloading a tar file, or use
mercurial:

\begin{verbatim}
hg clone http://bitbucket.org/darkskysims/darksky_tour
cd darksky_tour 
\end{verbatim}
There are a bunch of examples in here. Let's just walk through the first
one, which will create a nice visualization of all the particles within
a 50 Mpc box centered around the most massive galaxy cluster in ds14\_a.
The following is the \texttt{splat\_viz.py} example from the
darksky\_tour:

\begin{verbatim}
import yt
import numpy as np
from enhance import enhance
from yt.utilities.lib.image_utilities import add_rgba_points_to_image
from darksky_catalog import darksky

# Define a bounding box of 100 Mpc on a side.
center = np.array([-2505805.31114929,  -3517306.7572399, -1639170.70554688])
width = 50.0e3 # 5 Mpc
bbox = np.array([center-width/2, center+width/2])

ds = darksky['ds14_a'].load(midx=10, bounding_box=bbox)

ad = ds.all_data()
Npix = 1024
image = np.zeros([Npix, Npix, 4], dtype='float64')

cbx = yt.visualization.color_maps.mcm.RdBu
col_field = ad['particle_velocity_z']

# Calculate image coordinates ix and iy based on what your view width is
ix = (ad['particle_position_x'] - ds.domain_left_edge[0])/ds.domain_width[0]
iy = (ad['particle_position_y'] - ds.domain_left_edge[1])/ds.domain_width[1]

# Normalize the color field so that it doesn't get maxed out
col_field = (col_field - col_field.min()) / (col_field.mean() + 4*col_field.std() - col_field.min())
add_rgba_points_to_image(image, ix.astype('float64'), iy.astype('float64'), cbx(col_field))

# Write out a color-enhanced image
yt.write_bitmap(enhance(image), 'enhanced.png')
print 'Splatted %i particles' % ad['particle_position_x'].size
\end{verbatim}
This imports yt, a few extras from the darksky\_tour repository, and the
darksky\_catalog. After defining a bounding box into the entire dataset,
we load the data using a midx level 10 file. The yt dataset is returned
to the \texttt{ds} object, and we then manually splat particles onto a
canvas, colored by their velocity along the line of sight, and output an
image.

\begin{figure*}[htp]
\begin{center}
\dofig{
    \includegraphics[width=0.8\textwidth]{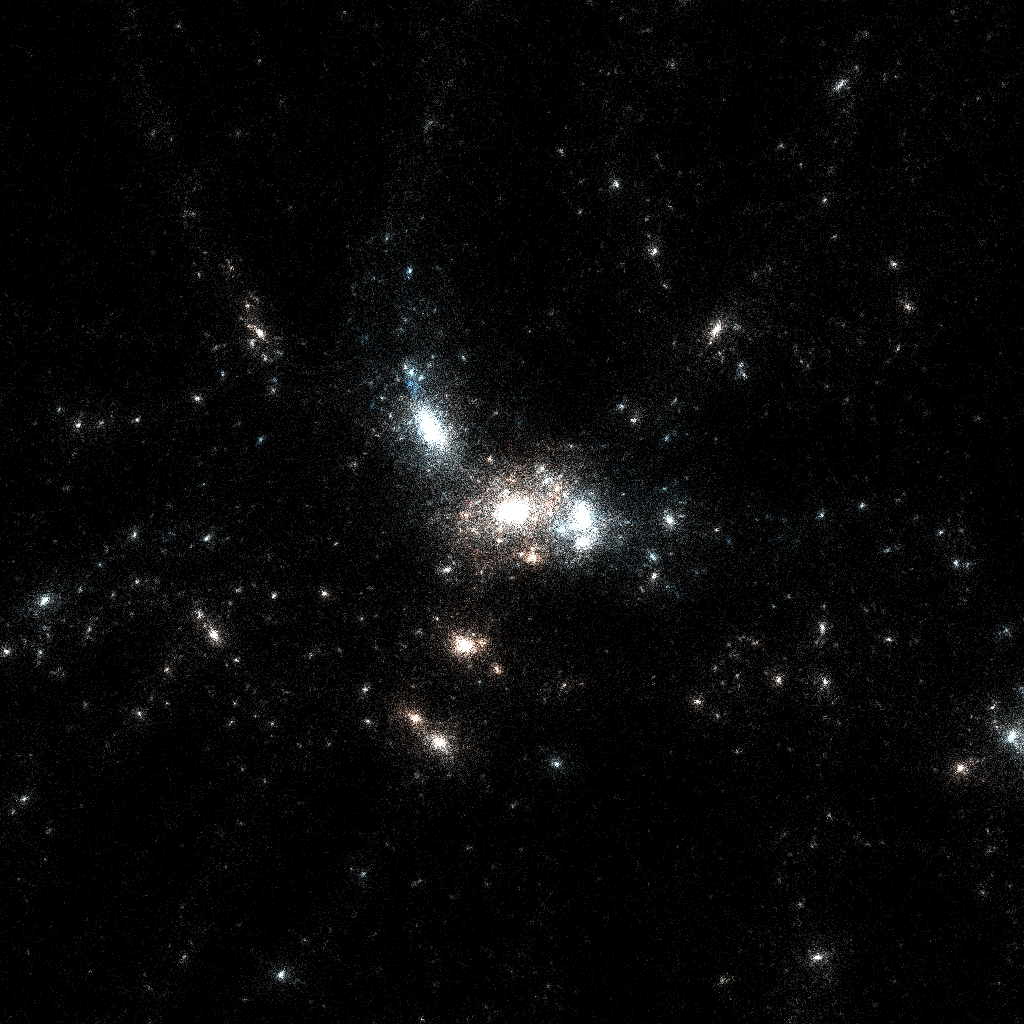}
}
\end{center}
\caption{Particles around the most massive galaxy cluster, colored
  by their line-of-sight velocity, generated by the 
  example in the Appendix. Particles are loaded over the WWW, then
  visualized locally.
}
\label{fig:lightcone}
\end{figure*}

} 

\end{document}